\documentclass[prr,floatfix,twocolumn,superscriptaddress,showpacs,preprintnumbers,longbibliography,nofootinbib]{revtex4-2}

\usepackage[utf8]{inputenc} 
\usepackage{bibunits} 
\usepackage{amssymb,amsfonts,amsmath}
\usepackage{float} 
\usepackage{amsthm}
\usepackage{enumitem}
\usepackage{bbold}
\usepackage[normalem]{ulem}
\usepackage{hyperref} 
\hypersetup{colorlinks=true, urlcolor=blue, citecolor=blue,linkcolor=blue}
\usepackage[all]{hypcap} 
\usepackage{graphicx}
\usepackage{dcolumn}
\usepackage{bm}
\usepackage{psfrag}
\usepackage{epsfig}
\usepackage{multirow}
\usepackage{color}
\usepackage{bbm}
\usepackage[FIGTOPCAP,raggedright,nooneline]{subfigure}
\usepackage{verbatim}
\usepackage{upgreek} 
\usepackage{physics}
\usepackage{todonotes}
\usepackage{tikz} 

\usepackage{orcidlink}

\newcommand{\ignore}[1]{}

\newcommand{\eq}{Eq.\,}
\newcommand{\eqs}{Eqs.\,}
\newcommand{\fig}{Fig.\,}

\newcommand{\cf} {cf.~}

\newcommand{\ie} {i.e.~}
\newcommand{\eg} {e.g.~}
\newcommand{\rref} {Ref.\,}
\newcommand{\rrefs} {Refs.\,}
 
\newcommand{\blue} {\textcolor{blue}}

\newcommand{\s}{\sigma}

\begin{document}

\defaultbibliography{refs_2}
\defaultbibliographystyle{apsrev4-2}

\begin{bibunit}[apsrev4-2-title]
	 
 \title{Multimode-cavity picture of non-Markovian waveguide QED}

\author{Dario Cilluffo\,\orcidlink{0000-0001-6862-0511}}
\affiliation{Institut f\"ur Theoretische Physik and IQST, Albert-Einstein-Allee 11, Universit\"at Ulm, 89069 Ulm, Germany}

\author{Luca Ferialdi\orcidlink{0000-0003-4016-3800}}
\affiliation{Universit$\grave{a}$  degli Studi di Palermo, Dipartimento di Fisica e Chimica -- Emilio Segr$\grave{e}$, via Archirafi 36, I-90123 Palermo, Italy}

\author{G. Massimo Palma\orcidlink{0000-0001-7009-4573}}
\affiliation{Universit$\grave{a}$  degli Studi di Palermo, Dipartimento di Fisica e Chimica -- Emilio Segr$\grave{e}$, via Archirafi 36, I-90123 Palermo, Italy}
\affiliation{NEST, Istituto Nanoscienze-CNR, Piazza S. Silvestro 12, 56127 Pisa, Italy}

\author{Giuseppe Calajò\orcidlink{0000-0002-5749-2224}}
\affiliation{Istituto Nazionale di Fisica Nucleare (INFN), Sezione di Padova, I-35131 Padova, Italy}

\author{Francesco Ciccarello\orcidlink{0000-0002-6061-1255}}
\affiliation{Universit$\grave{a}$  degli Studi di Palermo, Dipartimento di Fisica e Chimica -- Emilio Segr$\grave{e}$, via Archirafi 36, I-90123 Palermo, Italy}
\affiliation{NEST, Istituto Nanoscienze-CNR, Piazza S. Silvestro 12, 56127 Pisa, Italy}

\date{\today}

\begin{abstract}
We introduce a picture to describe and intrepret waveguide-QED problems in the non-Markovian regime of long photonic retardation times resulting in delayed coherent feedback. The framework is based on an intuitive spatial decomposition of the waveguide into blocks. Among these, the block directly coupled to the atoms embodies an effective lossy multimode cavity leaking into the rest of the waveguide, in turn embodying an effective white-noise bath. The dynamics can be approximated by retaining only a finite number of cavity modes { which} grows with the time delay. This description captures the atomic as well as the field's dynamics, even with many excitations, in both emission and scattering processes. As an application, we show that the recently identified non-Markovian steady states can be understood by retaining very few or even only one cavity modes.

\end{abstract}

\maketitle

Most quantum optics phenomena investigated so far occur under Markovian conditions{, i.e.,} lack of memory effects. One major reason behind this is that atom-photon interaction is usually weak, while light travels very fast. This allows to neglect photonic time delays/retardation times, which is a tremendous simplification of the dynamics underpinning standard tools such as the Lindblad master equation and the input-output formalism \cite{gardiner2004}. {However,} recent years have seen a growing attention to the {\it non-Markovian} regime of non-negligible photonic time delays, especially in the emerging area of waveguide Quantum ElectroDynamics (QED) which generally investigates the coherent interaction between quantum emitters and the one-dimensional (1D) field of a waveguide \cite{LiaoPhyScr16, RoyRMP17,Sheremet-RMP}. Such non-Markovian regime can today be accessed in some waveguide-QED experiments, \eg through superconducting qubits coupled to surface acoustic waves \cite{andersson2019non} or slow-light modes {near band edges} \cite{PainterPRX21} and even cold atoms coupled to fiber-ring resonators \cite{Arno-delay2023}. While complicating the dynamics considerably, time delays can be leveraged for a variety of unprecedented phenomena and applications, such as persistent quantum beats \cite{ZhengPRL13}, stabilization of Rabi oscillations \cite{CarmelePRL13,CarmelePRA13,GrimsmoPRL15}, peculiar inelastic two-photon scattering \cite{LaaksoPRL14,FangPRA15}, generation of photonic cluster states \cite{PichlerPNAS17}, excitation of dressed bound states in the continuum \cite{Calajo2019,TrivediPRA21,CarmeleBIC21}, enhanced Dicke superradiance \cite{SinhaPRL20, DincPRR19, SinhaPRA20,ZuecoPRA21}, anomalous population trapping \cite{CarmelePRR20}, stationary oscillations of giant atoms \cite{KockumPRR20}, enhanced energy-time entanglement \cite{CarmeleET21}, improved single-photon sources \cite{crowder2023improving}, genuinely non-Markovian steady states \cite{AskPRL22}.
\\
 
\begin{figure}
	\includegraphics[width=0.42 \textwidth]{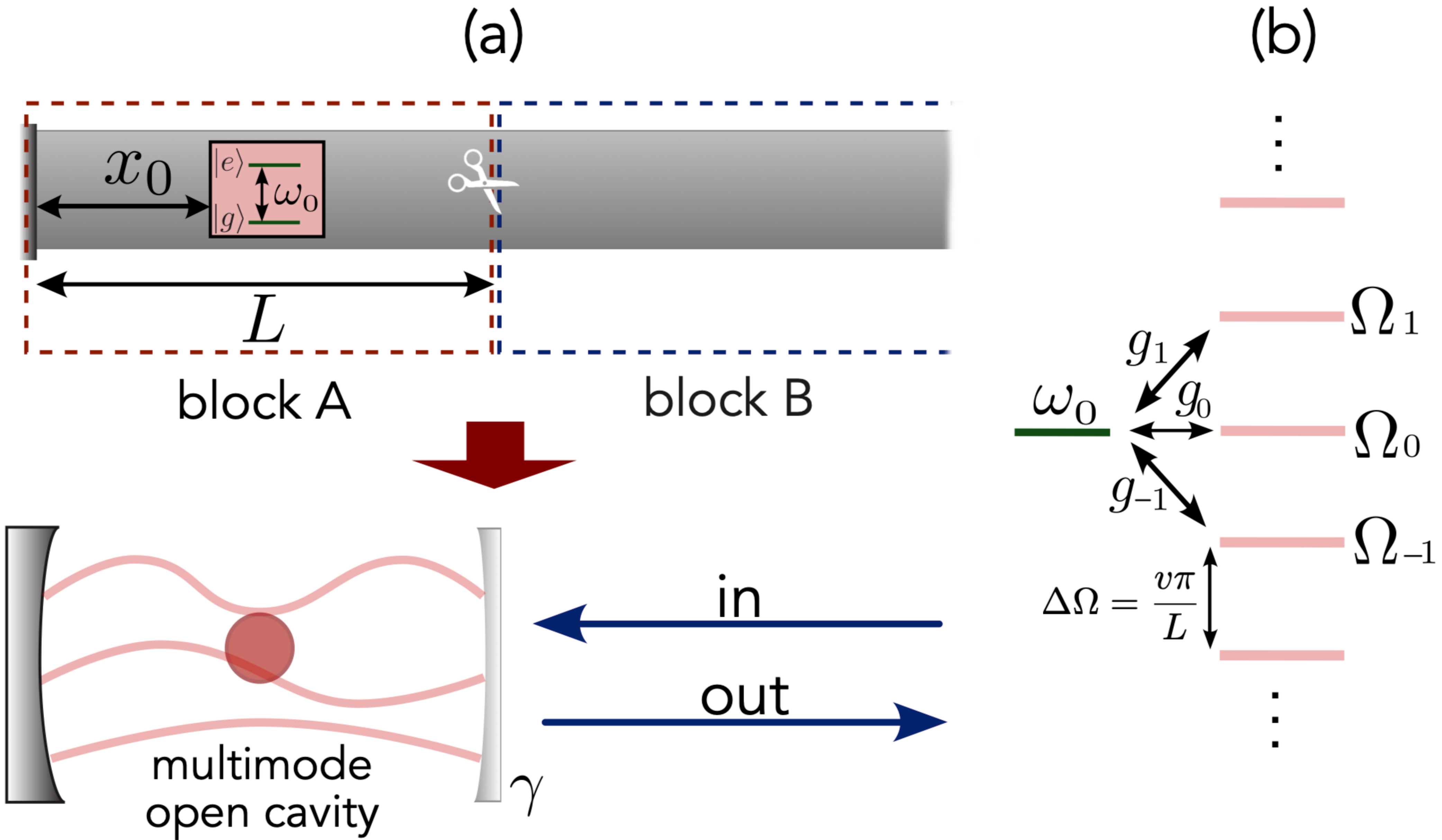}
	\caption{Basic setup and idea. (a) A two-level atom is coupled to a semi-infinite waveguide, whose left end acts like a perfect mirror. We conveniently decompose the waveguide into a pair of blocks $A$ and $B$ mutually coupled with rate $\gamma$. Block $A$ is an effective {\it multimode} and intrinsically {\it open} cavity {of length $L\gtrsim x_0$} in contact with block $B${, where the latter one works as a semi-infinite waveguide itself}. (b) Normal frequencies $\{\Omega_\nu\}$ of block $A$. The atom { of frequency $\omega_0$} resonates {with the cavity mode} of frequency $\Omega_0$ and is detuned from off-resonant modes $\Omega_{\pm 1}$, $\Omega_{\pm 2}$,...  The frequency spacing of $A$ modes $\Delta \Omega$ scales as $\sim 1/\tau$ with $\tau$ the time delay [\cf\eq\eqref{param0}], hence the stronger the non-Markovian effects, the larger is the number of $A$ modes to account for. Notice that each frequency $\Omega_\nu$ has width $\gamma\sim 1/\tau$ comparable with $\Delta \Omega$, reflecting the open nature of block $A$.}\label{fig-setup}
\end{figure}

From a merely computational viewpoint, it is well-established that such class of non-Markovian dynamics can be efficiently tackled via numerical methods \cite{pichlerPhotonic2016,RamosPRA16,trivediPRA19,WhalenPRA19,LeoCP19,dinc2019exact,CrowderPRA20,HughesPRR21,CrowderPRA22, zhang2022embedding,CarmelePRA22,vodenkova2023continuous}{; moreover diagrammatic approaches were developed \cite{DincPRA19,PletyuPRA21}}.
Notwithstanding, mostly due the daunting complication of {\it delayed} coherent feedback, the underlying physics remains generally involved and non-trivial to interpret. In particular, to our knowledge, no simple picture to understand such dynamics was so far identified even in the weakly non-Markovian limit{, i.e., the lowest non-trivial order in the characteristic time delay of the problem,} where one could expect a relatively simple effective model capturing the essential physics to exist.

With the above motivations, this work introduces a physical picture for describing and interpreting non-Markovian waveguide-QED problems that is built upon a real-space decomposition of the waveguide into blocks. The essential idea is inspired by cavity QED, where non-negligible time delays occur if the atom couples significantly to {\it many} cavity modes {(see \eg \rrefs \cite{MilonniPRA83,CookPRA87,gie1996cavity,RotterVolterra14})}: the longer the delay{, say the time taken by a photon to reach a cavity mirror,} the more cavity modes need to be considered. Of course, there is no actual cavity in a waveguide. {\it However,} nothing prevents one from viewing the latter as a set of communicating blocks [see blocks $A$ and $B$ in \fig\ref{fig-setup}(a)] and treating the {\it finite}-size block directly coupled to the atom [\ie block $A$ in \fig\ref{fig-setup}(a)] as an open cavity that leaks into the rest of the waveguide (this in turn described as a white-noise bath). If the size of the fictitious cavity [block $A$ in \fig\ref{fig-setup}(a)] is chosen to be comparable with the characteristic length of the problem, then the frequency spacing between block-$A$ normal modes scales as the inverse of the characteristic time delay $\tau$. 
Despite the relative arbitrariness of the block-$A$ boundaries, it turns out that this picture allows to define an effective Hamiltonian that fully captures {both spontaneous/driven} emission and photon scattering, including many-excitation dynamics where atomic non-linearities are important.

\section{ System and basic parameters} 
To present our theory, we will consider the case study where a two-level quantum emitter henceforth called ``atom" is coupled to a semi-infinite waveguide [see \fig\ref{fig-setup}(a)]. In spite of its apparent simplicity, this system hosts rich physics and is complex enough to show most salient effects of waveguide QED in the regime of long time delays (see \eg \rrefs\cite{DornerPRA02,TufarelliPRA13, TufarelliPRA14,GrimsmoPRL15,pichlerPhotonic2016,PichlerPNAS17,guimondDelayed2017,GoranPRA20,Calajo2019,zhang2022embedding,AskPRL22}), which includes (via a suitable mapping) emission of a giant atom \cite{GuoPRA17,andersson2019non} and even some paradigmatic sub- and super-radiance phenomena \cite{Gonzalez-BallesteroPRA16,SinhaPRL20}. The left end of the waveguide [see \fig\ref{fig-setup}(a)] works as a perfect mirror placed at distance $x_0$ from the atom. The waveguide sustains a 1D field with linear dispersion $\omega=v k$ {with $v$ the photon group velocity and $k$ the wave vector}. The atom's ground and excited states $|g\rangle$ and $|e\rangle$, respectively, are separated in frequency by $\omega_0=vk_0$; hence $k_0$ is the wave vector {modulus} of a photon resonant with the atom. {Assuming weak coupling, the Hamiltonian under the rotating-wave approximation} reads (we set $\hbar=1$) \cite{LiaoPhyScr16, RoyRMP17,Sheremet-RMP}

\begin{align}\label{Hmodel}
	H&=\omega_0\, \hat\sigma_+\hat\sigma_- \!-iv\!\!\int_{0}^{\infty}\!\! \!{\rm d} x\!\left[ \hat a^{\dagger}_R(x)\partial_x \hat a_R(x)\!-\! \hat a^{\dagger}_L(x)\partial_x \hat a_L(x)\right] \nonumber\\
	&\quad+ g\! \int_{0}^{\infty} \!\!\!\!{\rm d} x \left[ \hat\sigma_+\left( \hat a_{ L}(x)+ \hat a_{ R}(x)\right)\!+\!{\rm H.c.}\right]\delta(x{-}x_0)\,,\!\!\!
\end{align}
with $\partial_x=\frac{d}{d x}$, $ \hat\sigma_-=\hat\sigma_+^\dag=|g\rangle\langle e|$, 
$ \hat a_{R(L)}(x)$ the bosonic field operator annihilating a right-going (left-going) photon at position $x$ and $g$ the atom-photon coupling strength. 
This model is not analytically solvable in general, except for single-excitation dynamics {such as spontaneous emission or single-photon scattering} \cite{FangNJP18}.
The essential physical parameters are:
\begin{align}
	\Gamma=\frac{2g^2}{v}\,,\,\,\,\,\tau= \frac{2 x_0}{v}\,,\,\,\,\,\phi=2k_0x_0\,\,.\label{param0}
\end{align}
Here, $\Gamma$ is the standard decay rate that the atom would have without the feedback effect of the mirror{, i.e., as if the waveguide were infinite instead of ending at $x=0$}. Importantly, {here the {\it time delay} $\tau$} is the time taken by a photon resonant with the atom to travel {twice} the atom-mirror distance and  $\phi$ the corresponding accumulated phase. {The strength of non-Markovian effects is measured by $\Gamma \tau$, quantifying how long is the time delay $\tau$ compared to the lifetime $1/\Gamma$. }

\section{ Effective model}
We view the waveguide as two joint ``blocks" [see \fig\ref{fig-setup}(a)]: block $A${, corresponding to $x\in [0,L[$ with $L> x_0$,} and block $B${, corresponding to $x\ge L$}. Importantly, $A$ is the block directly coupled to the atom and has {\it finite} length. In contrast, block $B$ {is uncoupled from the atom and} has infinite length{. Note that block $B$} can be seen itself as a semi-infinite waveguide, {whose left edge lies at} $x=L$ [see \fig\ref{fig-setup}(a)]. Based on such block-decomposition of the waveguide, by calling $\hat\alpha_\nu$ and $\hat\beta_\omega$ respectively the normal-mode (bosonic) ladder operators of blocks $A$ and $B$, one can replace \eqref{Hmodel} with the effective Hamiltonian \cite{SM}
\begin{widetext}
	\begin{align}
		H_{\rm eff}=&\,\,\omega_0\, \hat\sigma_+\hat\sigma_-{+}\sum_\nu \Omega_{\nu}\, {\hat\alpha}^{\dagger}_{\nu}{\hat\alpha}_{\nu}{+}\int\!{\rm d}\omega\,\omega\, \hat\beta^\dag_\omega \hat\beta_\omega+ \sqrt{\frac{\gamma}{2\pi}}\,\sum_\nu\int\! {\rm d} \omega \,(  \hat\alpha_\nu^\dag  \hat\beta_\omega+{\rm H.c.})+\sum_\nu g_\nu \,( \hat\alpha_\nu^\dag \hat\sigma_-+{\rm H.c.})\label{Heff}\,,
	\end{align}
\end{widetext}
where 
\begin{align}
	\Omega_\nu&=\omega_0+v \frac{\nu\pi}{L}\,,\,\,\gamma=\frac{2v }{L}\,,\label{param1}\\
	g_\nu&=g\,(-1)^{\nu} \sqrt{\frac{2}{L}}\sin{\left(\frac{\nu\pi}{L} x_0+\frac{\phi}{2}\right)}\label{param2}\,
\end{align}
with $\nu$ running over all integers, while $\omega$ takes values throughout the real axis. 
The second (third) term of Hamiltonian \eqref{Heff} describes the free Hamiltonian of block $A$ (block $B$), where in particular $\Omega_\nu$ [\cf\eq\eqref{param1}] are the normal frequencies of block $A$ [see \fig\ref{fig-setup}(b)]. The fourth term of \eqref{Heff} couples block $A$ and block $B$ with a characteristic rate $\gamma$ given in \eq\eqref{param1}{. Notice that $1/\gamma$ is the characteristic time taken by a photon to leak out of block $A$}. Finally, the last term describes the coupling between the atom and each block-$A$ normal mode with corresponding coupling strength $g_\nu$ [see \fig\ref{fig-setup}(b)]. The expression \eqref{param2} of $g_\nu$ reflects the sinusoidal spatial shape of the $A$'s normal modes {just like a standard cavity-QED system}. As a hallmark of the present framework, $L$ (length of block $A$) is a free parameter of the model except for two conditions: {\it(i)} it must be strictly greater than $x_0$ to ensure that block $A$ contains the atom, but in practice is required to be still comparable with $x_0$ (more on this later on); {\it(ii)} $L$ must be a multiple integer of $\lambda_0/2$ with $\lambda_0=2\pi/k_0$ the atomic wavelength. Condition $(ii)$ makes sure that there is a block-$A$ mode resonant with the atom: this mode{, henceforth called ``resonant mode",} is labeled by $\nu=0$ [indeed \eq\eqref{param1} yields $\Omega_{\nu=0}=\omega_0$].

Hamiltonian $H_{\rm eff}$ formally describes an effective cavity-QED system in that the atom is coupled to a multimode lossy cavity (block $A$) leaking into a white-noise photonic bath (embodied by block $B$). Notice that block $A$ is an intrinsically {\it low-finesse} cavity: indeed [\cf\eqs\eqref{param1}] $\Delta \Omega=\Omega_{\nu+1}-\Omega_{\nu}\sim \gamma$, \ie the frequency spacing $\Delta\Omega$ between $A$ modes [\cf\fig\ref{fig-setup}(b)] is comparable with the loss rate $\gamma$ of block $A$. Physically, this stems from the inherently open nature of the fictitious cavity $A$ which fully lacks the mirror at $x=L$. 

To end up with \eq\eqref{Heff}, we resort to the standard discretization of a waveguide combined with weak coupling{, where the latter one allows to linearize the dispersion law} \cite{SM}. Discretizing the system this way enables a clean definition of the two blocks, leading to a natural identification of their associated Hamiltonian and normal modes, whose continuous limit is eventually worked out. {We notice that the possibility to express the electromagnetic field as a set of discrete modes defined in a finite region of space that are out-coupled to a continuum was shown in a general framework in \rrefs\cite{viviescas2003field,lentrodt2020ab}. }

\section{Dependence on time delay}

The essence of the present picture is that the dynamics of the joint system can be effectively described by replacing Hamiltonian \eqref{Hmodel} with \eqref{Heff}, where the latter can be approximated by retaining only a finite number of $A$ modes {which however} eventually grows with the time delay $\tau$. To see the last key property, recall that we require $L$ to be of the order of $x_0$. Thus the spacing of $A$ modes $\Delta\Omega=\pi v/L$ is of the order of $ \tau^{-1}$ [see \eqs\eqref{param0}, \eqref{param1} and \fig\eqref{fig-setup}(b)]{, i.e., the detuning between the atom and off-resonant modes $|\nu|\ge 1$,} scales as the inverse of time delay (recall that block $A$ is defined so as to ensure $\omega_{0}=\Omega_0$).
Accordingly, in the Markovian regime of vanishing $\tau$ {all these off-resonant modes are far-detuned from the atom so that only the resonant mode} needs to be accounted for [\cf\fig\ref{fig-setup}(b)]. 

As the time delay grows up so that non-Markovian effects get increasingly important, more and more off-resonant modes must be retained in general. In other words, for given $\tau$, one neeeds to retain all the modes  $\nu=0,\,\pm 1,...,\pm N_A$ with $N_A$ eventually growing with $\tau$. Thus the {\it multimode} nature of block $A${, instead of a more canonical one-mode cavity,} reflects occurrence of retardation effects: while this is a well-known fact in standard cavity QED \cite{MilonniPRA83,CookPRA87,RotterVolterra14}, the present framework provides ground to take advantage of this property also in the study of {\it waveguide}-QED systems.

\section{Testing the framework}
\begin{figure}
	\includegraphics[width=0.48 \textwidth]{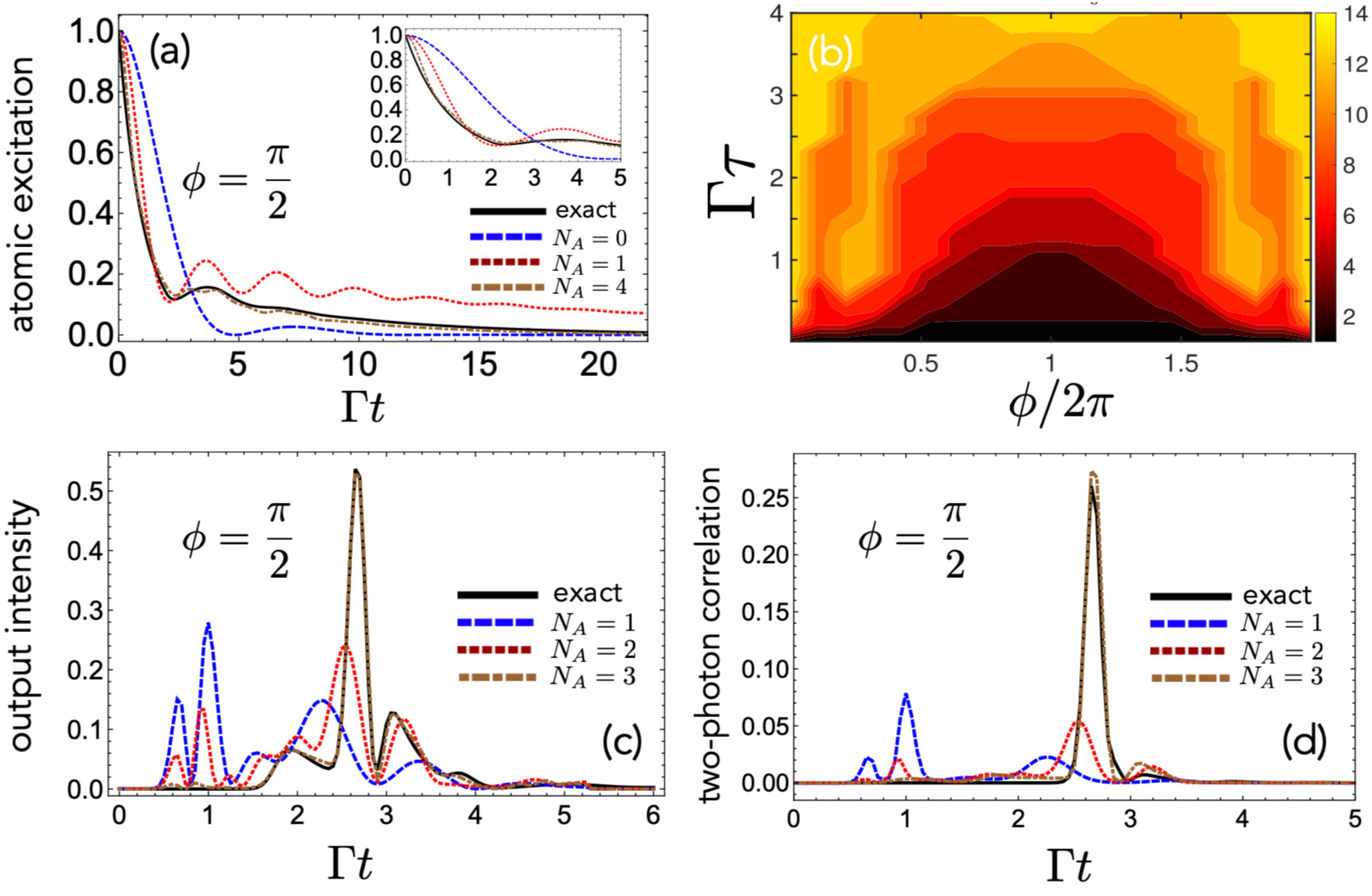}
	\caption{(a) Atomic excitation, i.e.,  $\langle e| \rho|e\rangle=\rho_{ee}$ with $\rho$ the atom's density matrix, when the atom is initially in state $\ket{e}$ and the field in the vacuum state: exact analytical solution \cite{DornerPRA02,TufarelliPRA13} (black line) versus the approximated one using \eq\eqref{Heff} by retaining only modes $\nu=0, \pm1,...,\pm N_A$ for $\phi=\pi/2$ and $\Gamma \tau=2$. {(b) Dependence on $\phi$ and $\Gamma\tau$ of the minimum number of modes $\tilde{N}_A$ yielding a mean square root deviation below $1\%$ between the approximate and exact solutions of $\rho_{ee}(t)$ for the same process as in panel (a).} (c)-(d) Field's output intensity  (c) and two-photon correlation function  (d)  after the scattering of a left-incoming Gaussian coherent wave packet (see \rref\cite{SM}). We set the wave packet's width to $W=2.5\Gamma$ and the average number of photons to $n_{\rm ph}=0.5$ with phase $\phi=\pi/2$ and delay $\Gamma \tau=4$. In (a)-(d), we set $L=2 x_0$.}\label{fig-test}
\end{figure}

{To check the effectiveness of the waveguide decomposition into blocks, in \fig\ref{fig-test}(a), (c) and (d) we set $L=2 x_0$, $\Gamma \tau\ge2$ (relatively long delay) 
and the representative phase $\phi=\pi/2$ 
for two paradigmatic dynamics: spontaneous emission [panel (a)] and scattering of a coherent-state wave packet [(c)-(d)]. }

In each case, the dynamics predicted by the effective model in \eq\eqref{Heff} by retaining only the $A$ modes $\nu=0, \pm1,...,\pm N_A$ is compared with the exact solution of \eqref{Hmodel} obtained through either analytical methods when available [as in the single-excitation dynamics of {panel (a)}] or numerical simulations based on Matrix Product States \cite{pichlerPhotonic2016} [panels (c)-(d)].
As $N_A$ is made larger, the mismatch between exact and approximated dynamics gets smaller and smaller until becoming negligible. Notice that, the dynamics in panels (c)-(d) involves many excitations, providing evidence that the picture is effective even when the atom's intrinsic nonlinearity has substantial effects. This is especially striking in panel \fig\ref{fig-test}(d), reporting the two-photon correlation function of scattered light, which clearly shows a multi-photon peak {that adds to the delayed single-photon peak of the output intensity in panel (c)}. {Analogous conclusions hold for different settings of the parameters, including the special value of phase $\phi=2\pi$ where it is known that the atom does not fully decay \cite{DornerPRA02,TufarelliPRA13}.}

We note that the convergence rate is generally dependent {on the set values of the relevant parameters},
a major reason being the sinusoidal dipendence on these parameters of the coupling strength $g_\nu$ [\cf\eq\eqref{param2}]. 
For analogous reasons, in general the convergence is not strictly monotonic, i.e., it can happen that $N_A+1$ modes perform as $N_A$ or even worse \cite{SM}. Notwithstanding, convergence eventually occurs because $|g_\nu|\le g \sqrt{2/L}$ for any $\nu$ while the detuning $|\Omega_\nu-\omega_0|$ grows linearly with $\nu$ [\cf\eqs\eqref{param1}-\eqref{param2}]. {Evidence of this is provided in \fig\ref{fig-test}(b), where we study the dependence on the phase $\phi$ and rescaled time delay $\Gamma \tau$ for $L=2 x_0$ of the required number of modes to capture the exact dynamics  for the paradigmatic process of spontaneous emission. Specifically, we plot the minimum number of modes $\tilde{N}_A$ yielding a mean square root deviation below $1\%$ between the approximate and exact solutions (see \rref\cite{SM} for details). As expected, $\tilde{N}_A$ eventually grows with $\Gamma\tau$, witnessing that, as non-Markovian effects get stronger, more and more modes of block $A$ generally need to be accounted for in the description. Interestingly, a fast (low) convergence rate as a function $N_A$ occurs for $\phi\sim \pi$ ($\phi\sim 0,2\pi$), at which values the considered dynamics exhibits the weakest (strongest) non-Markovian behaviour as measured by a rigorous non-Markovianity measure \cite{TufarelliPRA14}.}

Although somewhat implicit in the above, it is worth stressing here that the advantage of the framework is not to provide a fast computational numerical tool (where efficient techniques exist already) but rather an intuitive physical picture connecting non-Markovian waveguide QED with cavity QED. This is illustrated next with some important instances.

\section{Markovian limit and Purcell effect}
For $\Gamma\tau\ll 1$ we are in the Markovian regime: time delay is negligible, but the feedback provided by the mirror affects atomic emission resulting in a decay rate modulated by $\phi$ as $\Gamma'(\phi)=2\Gamma \sin^2\frac{\phi}{2}$ \cite{DornerPRA02,TufarelliPRA13}, which was experimentally confirmed  \cite{HoiNatPhy15}.
Thus emission can be either enhanced or suppressed; in particular $\Gamma'=2\Gamma$ for $\phi=(2m+1) \pi$ while $\Gamma'=0$ for $\phi=2m \pi${, where $m$ is an integer}. Now, using our framework, we see that, due to $\Delta\Omega\sim 1/\tau$ [\cf\eq\eqref{param2}], for negligible $\tau$ the off-resonant modes of block $A$ are very far-detuned from the atom and thus can be neglected. {Only the resonant mode} therefore needs to be accounted for. Its bandwidth is, {however}, very large since we also have $\gamma\sim \Delta\Omega \sim 1/\tau$ [\cf\eq\eqref{param2}]. The system therefore reduces to an atom coupled to a standard one-mode cavity, but in the {\it bad cavity} limit. The corresponding atom-mode coupling strength is  $g_0=g\sqrt{\frac{2}{L}}\sin{\frac{\phi}{2}}$ [\cf\eq\eqref{param2}], which indeed vanishes for $\phi=2m\pi$ meaning that in this case the atom sits right on a cavity field's node and is thereby unable to emit. Standard cavity-QED theory in the bad-cavity limit then predicts the decay rate $4g_0^2/\gamma$, which indeed exactly matches $\Gamma'(\phi)$. This shows that the action of the mirror, despite no actual cavity is present, can still be seen as a manifestation of the standard Purcell effect in a cavity.
\begin{figure}
	\includegraphics[width=0.4 \textwidth]{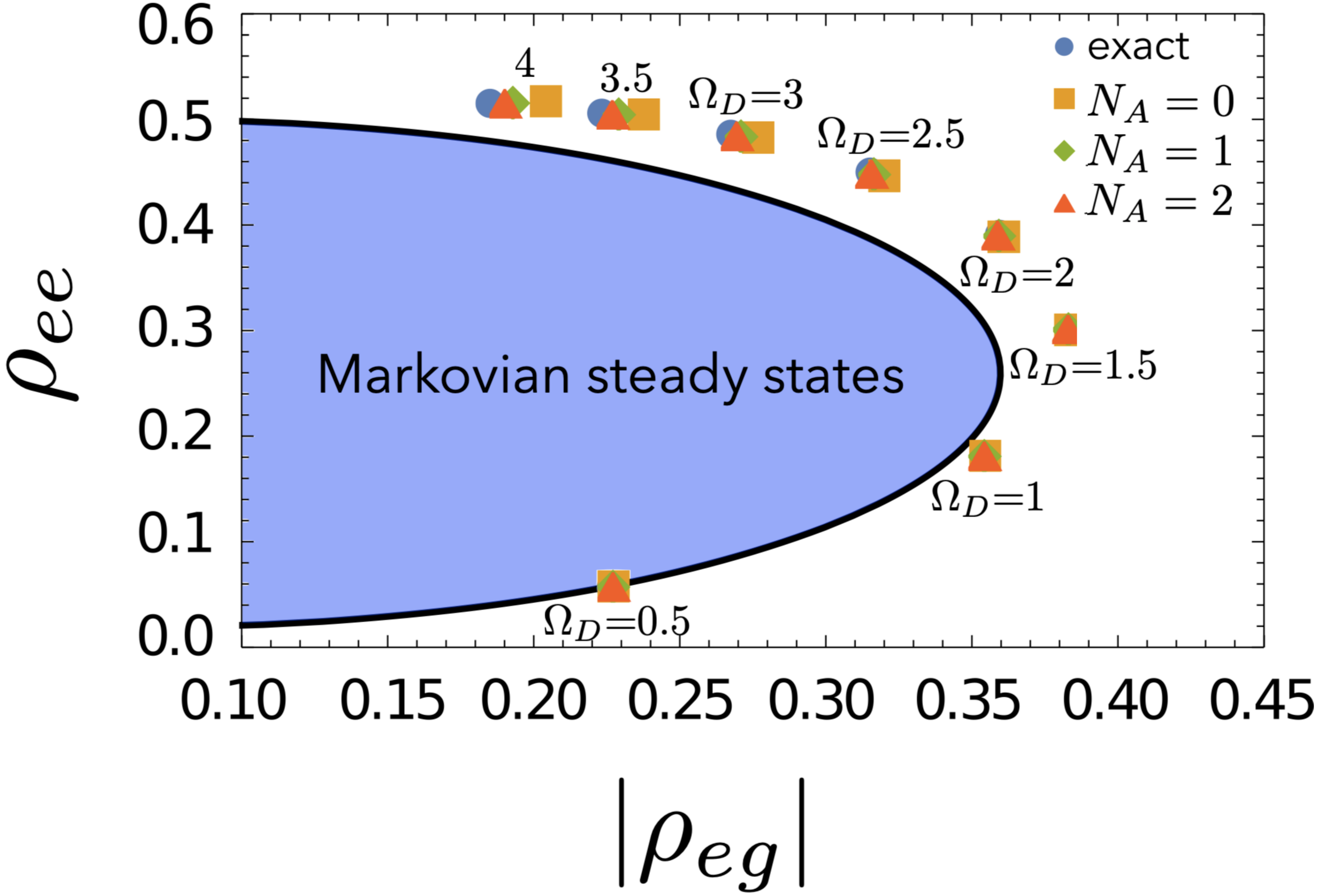}
	\caption{Non-Markovian steady states. {Each point $(|\rho_{eg}|,\rho_{ee})$, with $\rho_{eg}$ ($\rho_{ee}$) the coherences (excited-state population), fully specifies a possible steady state of a driven atom emitting into the semi-infinite waveguide.} Emission into a Markovian bath [see \eq\eqref{ME} for {$\dot\rho=0$}] can only yield steady states within the shaded area \cite{AskPRA19}. The blue dots are exact steady states computed through MPS simulations for different Rabi frequencies $\Omega_D$ (in units of $\Gamma$) 
    and for $\Gamma \tau=0.25$ and $\phi=\pi$. The other points (squares, rhombi and triangles) {are steady states} calculated through the effective Hamiltonian \eqref{Heff} by retaining {only} $N_A$ modes of block $A$ for $L=x_0$. {Each non-Markovian steady state is computed by evolving the whole system for a sufficiently long time and tracing over the field/block modes until the atom's reduced state -- in a frame rotating at frequency $\Omega_D$ -- reaches an asymptotic value $\rho$. The corresponding values of $|\rho_{eg}|$ and $\rho_{ee}$ then define a point on the diagram.}  }\label{ellipse}
\end{figure}

\section{ Non-Markovian steady states}
Recently, Ask and Johansson showed that a driven atom in front of a mirror for non-negligible $\Gamma\tau$ can reach steady states unattainable in a standard Markovian bath \cite{AskPRL22}. Specifically, let
{\begin{equation}
	\dot{\hat\rho} = -i\left[\tfrac{1 }{2}\Omega_D\hat\s_x , \hat\rho\right] +\kappa \,\mathcal{D}[\hat\sigma_-]\hat\rho  + \kappa_\phi \,\mathcal{D}[\hat{\sigma}_+ \hat{\sigma}_-] \label{ME}
\end{equation}}
with $\mathcal{D}[A]\hat\rho=A\hat\rho A^\dag-\tfrac{1}{2}\{A^\dag A,\hat\rho\}$, be the standard Markovian master equation (ME) of an atom subject to a classical drive of Rabi frequency $\Omega_D$, pure dephasing with rate $\kappa_\phi$ and decaying into a Markovian bath with rate $\kappa$. An atom's state $\rho$ is fully specified by the {excited-state} population $\rho_{ee}$ and coherence $\rho_{eg}$. After a transient, \blue{the }emitter reaches a steady state obtained by imposing $\dot{\rho}=0$ in \eq\eqref{ME}. It turns out that, irrespective of \blue{$\Omega_D$} and $\kappa$, the steady state must lie within the shaded region bounded by the elliptical line shown in \fig\ref{ellipse}. This bound in particular holds for a driven atom in a semi-infinite waveguide in the Markovian regime $\Gamma\tau\ll1$ (see previous section). {However,} it can be violated in the non-Markovian regime: see, e.g., the blue dots in \fig\ref{ellipse} computed via exact MPS numerical simulations \cite{AskPRL22,SM} corresponding to the steady states occurring for $\Gamma \tau=0.25$, $\phi=\pi$ and (from bottom to top, see curved arrow) growing values of $\Omega_D$. 

The steady states in \fig\ref{ellipse} can be well-approximated using our block-decomposition framework, as shown by \fig\ref{ellipse} where the agreement with the exact solution (blue dots) grows with  $N_A$. Remarkably, retaining even a {\it single} cavity mode ($N_A=0$; orange squares) provides an excellent quantitative approximation at low values of $\Omega_L/\Gamma$ and, as long as population $\rho_{ee}$ is concerned, even at larger ones; in any case, on a qualitative ground, it appears to capture most of the relevant physics. In this case, the emitter's steady state can be obtained through a partial trace from the bipartite Markovian ME governing the evolution of $\hat\varrho${, i.e., the joint state of the atom and mode $\nu=0$,}
	\begin{align}
		\dot{\hat\varrho} = -i[\tfrac{1}{2}\Omega_D\hat\s_x + g_0 (\hat\alpha^\dag_0 \hat\s_- + {\rm H.c.}), \hat\varrho]
		+\gamma \mathcal{D}[\hat\alpha_0]\hat\varrho \,,
	\end{align}
with $\gamma$ and $g_0$ given by \eqs\eqref{param1} and \eqref{param2}. We point out that the occurrence of population inversion at long times, i.e., $\rho_{ee}{>}1/2$, is a sufficient condition to reach non-Markovian steady states beyond the elliptical bound in \fig\ref{ellipse}. Occurrence of population inversion for a driven two-level system coupled to a lossy cavity mode is a well-established quantum optics effect  \cite{lindberg1988steady}, which highlights a further interesting connection between non-Markovian waveguide QED and cavity QED.

\section{ Conclusions}
We presented a picture to describe and interpret waveguide-QED dynamics with delayed feedback. After decomposing the waveguide into blocks, the block coupled to the atom is viewed as an open cavity leaking into the rest of the waveguide. The longer the time delay, the more modes of such open cavity generally need to be accounted for. The picture captures both the atom and field dynamics, even when many excitations are present.

{ While we focused on one atom in a semi-infinite waveguide, the framework can be  extended to many emitters. A possible modular generalization is discussed in Appendix \ref{app:E} and illustrated in the representative instance of two atoms coupled to an infinite waveguide by decomposing this into four adjacent open blocks, where each of two central blocks contains one atom. A natural generalization of \eq\eqref{Heff} provides the effective Hamiltonian, which can likewise be approximated by retaining a finite number of modes per central block and treating the pair of outer blocks as Markovian baths. The method was successfully tested by demonstrating its ability to capture non-Markovian super- and sub-radiance \cite{SinhaPRL20}.}

The idea that atoms in waveguides could be understood in terms of effective cavities appeared several times {(see \eg \rrefs\cite{ChangNJP12,GuimondPRA16,AskPRA19,GoranPRR21,Hughes3qubits})}, sometimes relying on the well-known mirror-like behavior of an atom \cite{Shen2005}. In contrast, the cavity central to our framework relies on a fully transparent fictitious mirror. {However,} it is instrumental to the establishment of a sharp link between emission phenomena featuring delayed feedback and cavity-QED physics with the bonus that the picture can capture the waveguide-field dynamics as well. 
Our theory can be seen to define a {so called} Markovian embedding or dilation in the following sense: one replaces an open system{ -- the atom in our case --} immersed in a non-Markovian bath with an enlarged open system {-- the atom plus $A$ modes here --} that is instead immersed in a Markovian bath. This is arguably the most common strategy to attack non-Markovian problems and is typically accomplished by adding auxiliary lossy modes to the open system \cite{TamaPRL18,CampbellPRA18,TrivediPRL21,zhang2022embedding,TamaPRL22}. Two remarkable features{, however,} stand out in the present approach: (i) the open-cavity modes have a clear physical meaning and can be straightforwardly visualized as degrees of freedom taken out of the bath (\ie the waveguide); (ii) besides the open system, the framework can describe as well the bath dynamics. 
We anticipate that this work could offer a new alternative approach to understanding and interpreting waveguide-QED phenomena in the relatively unexplored non-Markovian regime.
\\
\\
\begin{acknowledgments}
\indent D.C.~acknowledges support from the BMBF project PhoQuant (grant no.~13N16110) and the state of Baden-Württemberg through bwHPC and the German Research Foundation (DFG) through grant no.~INST
40/575-1 FUGG (JUSTUS 2 cluster). G.M.P.~acknowledges support from MUR under PRIN Project no.~2022FEXLYB
``Quantum Reservoir Computing" (QuReCo).
G.C.~acknowledges that results incorporated in this standard have received funding from the  T-NiSQ consortium agreement financed by QUANTERA 2021 and by the Italian  Ministry of University and Research MUR Departments of Excellence grant 2023-2027 "Quantum Frontiers" (FQ). Numerical simulations were performed using QuTip \cite{JOHANSSON20121760} and mpnum \cite{suess2017mpnum}. 
F.C.~acknowledges support from European Union-Next Generation EU through projects: Eurostart 2022 ``Topological atom-photon interactions for quantum technologies"; PRIN 2022--PNRR no.~P202253RLY ``Harnessing topological phases for quantum technologies"; THENCE--Partenariato Esteso NQSTI--PE00000023--Spoke 2 ``Taming and harnessing decoherence in complex networks".
The authors would like to thank D. Lentrodt for fruitful discussions.
\end{acknowledgments}

		\putbib	
\end{bibunit}
	
		\clearpage
					
		
\begin{bibunit}[apsrev4-2-title]

				\onecolumngrid
%
			
\appendix
    	\section{Derivation of the effective Hamiltonian }
	
	Here we derive the effective Hamiltonian \eqref{Heff} by first discretizing the waveguide, then decomposing it into blocks and finally taking the continuous limit. We start with a review of the discretization of a continuos waveguide.
	
	\subsection{Discretized waveguide: review}
	
	Consider a homogeneous coupled-cavity array with cavities located at positions $n$, whose free Hamiltonian has the usual tight-binding form
	\begin{equation}
	\hat 	H_F=\omega_c\sum_n 	\hat  a_n^\dag  	\hat a_n-J \sum_n  (	\hat  a_{n+1}^\dag  	\hat  a_n+{\rm H.c.})\,\label{HB-wav}\, ,
	\end{equation}
	with $\omega_c$ the frequency of each cavity and $J$ the cavity-cavity hopping rate.
	
	Under periodic boundary conditions ($	\hat  a_1\equiv 	\hat  a_{N+1}$) the waveguide's normal modes  are the usual plane waves with dispersion law 
	\begin{equation}
		\omega_k=\omega_{c}-2J\cos k\label{wk2}\, ,
	\end{equation}
   which gives rise to an energy band of width $4J$, where $k_m=2\pi m/ (N+1)$ with $m=1,...,N$. The corresponding group velocity is $v_k=d \omega/dk=2J  \sin k $, hence (far from the band edges)  the dispersion law can be linearized around a specific wavevector $k_0$ as $\omega_k\simeq (\omega_{c}-2J\cos k_0)+v (k-k_0) $ with $v=v_{k_0}$. 
   
   In the presence of an atom coupled to the waveguide at position $n_0$, the total Hamiltonian reads
   	\begin{equation}
   	\hat 	H=\omega_{0}\hat \sigma_{+}\hat \sigma_-+\hat H_F+\hat V_{AF} \,.\label{Htot}
   \end{equation}
   with the atom-field interaction Hamiltonian given by
   	\begin{equation}
   \hat V_{AF}=g\, (\hat \sigma_{+} \hat a_{n_0}+{\rm H.c.})\label{vaf}\,\,.
  \end{equation}
   Note that the coupling-strength $g$ here is defined differently from the one in Eq. \eqref{Hmodel} of the main text, and indeed they have different dimensions.
   If the atom is tuned inside the photonic band (\ie $\omega_{c}-2 J<\omega_{0}< \omega_{c}+2 J$), only photonic modes of frequency $\omega_k$ close to $\omega_0$ need to be considered, hence one can linearize the dispersion law around $\pm k_0$ with $k_0$ defined by $\omega_0=\omega_{k_0}$.
   It can be shown that the real-space representation of the linearized field's Hamiltonian so obtained is analogous to the second term of \eqref{Hmodel}. 
   
If we now consider {\it open} boundary conditions, such that $	\hat  a_0=	\hat  a_{N+1}=0$, the spectrum \eqref{wk2} is unaffected but the normal modes now have sinusoidal shape
\begin{equation}\label{alpha1}
    \hat{\alpha}_m=\sqrt{\frac{2}{N+1}}\sum_{n =1}^{N}\,\!\sin{\left(k_m n \right)} \,\hat{a}_n \,\,\,\,{\rm with}\,\,\,\,k_m=\frac{\pi m}{N+1}\,\,\,{\rm for} \,\,m=1,...,N\,.
\end{equation}
	In the limit $N\rightarrow \infty$ we obtain a semi-infinite waveguide (or equivalently a waveguide subject to a hard-wall boundary condition at $n=0$).

	\subsection{Decomposition into blocks}

	\begin{figure}[t]
		\centering
		\includegraphics[width=0.85\textwidth]{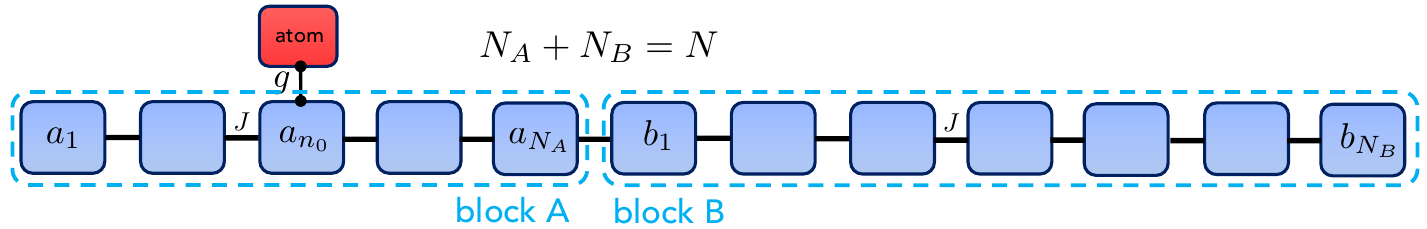}
		\caption{Decomposition of the discretized waveguide into blocks $A$ and $B$ with $b_1\equiv a_{N_A+1}, ...,b_{N_B}\equiv a_{N}$. Note that $N_A\ge n_0$, meaning that block $A$ is directly coupled to the atom.\label{fig-blocks-sm}}
	\end{figure}
	
	We start from Hamiltonian~\eqref{HB-wav} under open boundary conditions and split the discretized waveguide in two blocks (see \fig\ref{fig-blocks-sm}): block $A$ for $1\le n\leq N_A$ with $N_A\ge n_0$ and block B for $N_A+1\leq n\leq N$ (so that block $A$ is the one of the two directly coupled to the atom). Corresponding to this block decomposition, we rearrange the free field's Hamiltonian as
	\begin{align}
		\hat H_F=\underbrace{\omega_c \sum_{n=1}^{N_A} \hat a^\dag_n\hat a_n-J \sum_{n =1}^{N_A-1} \left(\hat a_n  \hat a_{n +1}^\dag+{\rm H.c.}\right)}_{\hat H_A}+\underbrace{\omega_c \sum_{n=1}^{N_B} \hat b^\dag_n\hat b_n-J \sum_{n =1}^{N_B-1} \left(\hat b_n  \hat b_{n +1}^\dag+{\rm H.c.}\right)}_{\hat H_B}\,\,\underbrace{-J \left(\hat a_{N_A} \hat b_{1}^\dag+{\rm H.c.}\right)}_{\hat V_{AB}}\label{Hdec}
	\end{align}
	where $N_B=N-N_A$ is the size of block $B$. Notice that for convenience we have re-named the bosonic site modes of block $B$ as $\hat b_n{=}\hat a_{n-N_A}$. In \eq\eqref{Hdec}, $\hat H_A$ ($\hat H_B$) is the free Hamiltonian of block $A$ (block $B$). It is important to stress that the two blocks are mutually {\it coupled} with interaction Hamiltonian $\hat V_{AB}$ and coupling strength just equal to the photon hopping rate $J$: their coupling guarantees a photon coming from block $A$ to reach $B$ (or the other way around) without suffering any back-reflection.

	The sine-shaped normal modes of block $A$ read [compare with \eq\eqref{alpha1}] 
    \begin{equation}
      \hat{\alpha}_m =\sqrt{\frac{2}{N_A+1}}\sum_{n =1}^{N_A}\,\!\sin{\left(k^A_m n \right)} \,\hat{a}_n \,\,\,\,\,\,{\rm with}\,\,\,\,k^A_m=\frac{\pi m}{N_A+1}\,\,\,{\rm for} \,\,m=1,...,N_A\,.
    \end{equation}
	In terms of these, the free Hamiltonian of block $A$ takes the diagonal form
	\begin{equation} \label{H0f-diagonal}
		\hat{H}_A\!=\!\sum_{m=1}^{N_A} \omega^A_{m}\, \hat{\alpha}^{\dagger}_{m}\hat{\alpha}_{m}\,\,,
	\end{equation}
	where $\omega^A_m=\omega_c-2J \cos{k^A_m}$.
	Real-space block-A operators can be expressed in terms of these normal modes as
	\begin{equation}\label{nba}
	    \hat a_n \!=\sqrt{\frac{2}{N_A+1}}\,\sum_{m=1}^{N_A}\,\!\sin{(k^A_m n )} \,\hat{\alpha}_m\,.
	\end{equation}
	 This allows us to express even the atom-field interaction Hamiltonian \eqref{vaf} in terms of the $A$'s normal modes as
	\begin{align}
		\hat V_{AF}&= \sum_{m=1}^{N_A} g_m\left( \hat\sigma_+\hat \alpha_m+{\rm H.c.}\right)\,,
	\end{align}
	where $g_m$ measures how strongly is the atom coupled to the $m$-th normal mode of block $A$
	\begin{align}\label{gm}
		g_m=g\, \sqrt{\frac{2}{N_A+1}}\sin{\left(k^A_m{n_0} \right)}\,.
	\end{align}
	Likewise, using the same decomposition \eqref{nba} for $n=N_A$, the coupling Hamiltonian between blocks A and B [\cf\eq\eqref{Hdec}] is arranged as
	\begin{equation}
	\hat V_{AB}=-J\sqrt{\frac{2}{N_A+1}}\sum_{m=1}^{N_A}\sin{\left(k^A_m N_A\right)} \,\hat{\alpha}_{m}\,\hat b_1^\dag=\sum_{m=1}^{N_A} \xi_m\,\left(\hat \alpha_m \hat b_1^\dag+{\rm H.c.}\right)\,,\label{vab1}
	\end{equation}
	where we defined
	\begin{align}
		\xi_m=J\sqrt{\frac{2}{N_A+1}}\,(-1)^{m}\sin\left(\frac{m\pi }{N_A+1}\right)\,,\label{xim1}
	\end{align}
	and used the identity
	\begin{align}
		\sin{\left(k^A_m \,N_A\right)}= -(-1)^m\sin\left(k^A_m\right)\,.
	\end{align}
	\\
	\\
	Analogously to block $A$, we can define normal modes also for block $B$ as [\cf\eq\eqref{nba}]
	\begin{equation}
	    \hat{\beta}_m\!=\sqrt{\frac{2}{N_B+1}}\,\,\sum_{n =1}^{N_B}\,\!\sin{\left(k^B_m n \right)} \,\hat{b}_n\,\,\,\,{\rm with}\,\,\,\,k^B_m=\tfrac{\pi m}{N_B+1}\,\,\,{\rm for} \,\,m=1,...,N_B
	\end{equation}
	such that [\cf\eq\eqref{H0f-diagonal}] $\hat{H}_B\!=\!\sum_{m=1}^{N_A} \omega^B_{m}\, \hat{\beta}^{\dagger}_{m}\hat{\beta}_{m}$ with $\omega^B_m=\omega_c-2J \cos{k^B_m}$. The inverse transformation reads
	\begin{equation}
	\hat b_n \!=\sqrt{\frac{2}{N_B+1}}\,\sum_{m=1}^{N_B}\,\!\sin{\left(k^B_m n \right)} \,\hat{\beta}_m\,,
	\end{equation}
	and allows now to arrange the $A$-$B$ interaction Hamiltonian \eqref{vab1} as
	\begin{equation}
	\hat V_{AB}=\sum_{m=1}^{N_A}\sum_{m'=1}^{N_B} \xi_m\chi_{m'}\,\left(\hat \alpha_m \hat \beta_{m'}^\dag+{\rm H.c.}\right)\,.
	\end{equation}
	with
	\begin{align}
		\chi_{m}=\sqrt{\frac{2}{N_B+1}}\sin{\left( \frac{m\pi}{N_B+1}  \right)}
	\end{align}
	\\
	\\
	To summarize, in terms of block-$A$ and block-$B$ normal modes the total Hamiltonian reads
	\begin{align}
		\hat H= \omega_{0}\hat \sigma_{+}\hat \sigma_-+\underbrace{\sum_{m=1}^{N_A} \omega^A_m\, \hat{\alpha}^{\dagger}_{m}\hat{\alpha}_{m}+\sum_{m=1}^{N_B} \omega^B_m\, \hat{\beta}^{\dagger}_{m}\hat{\beta}_{m}+\sum_{m=1}^{N_A}\sum_{m'=1}^{N_B} \xi_m\chi_{m'}\,\left(\hat \alpha_m \hat \beta_{m'}^\dag+{\rm H.c.}\right)}_{=\,\hat H_F}+\underbrace{\sum_{m=1}^{N_A} g_m\left( \hat\sigma_+\hat \alpha_m+{\rm H.c.}\right)}_{=\,\hat V_{AF}}\label{Ht}
	\end{align}
	where we recall that
	\begin{align}\label{kkp}
		\omega^A_m=\omega_c-2J \cos{k^A_m}\,,\quad\omega^B_m=\omega_c-2J \cos{k^B_m}\,.
	\end{align}

	\subsection{Continuous limit and linearization}
	
	Since we are considering a semi-infinite waveguide, the length of block B must diverge as $N_B\rightarrow \infty$. Accordingly, block-$B$ normal ladder operators $\{{\hat\beta}_m\}$ become a continuum of singular bosonic modes $\{\hat\beta(k)\}$ with $0\le k <\pi$ fulfilling $[\hat\beta(k),\hat\beta^\dag(k')]=\delta(k-k')$. Specifically, $\hat \beta(k)$ is obtained as the continuous limit of the rescaled ladder operators $\hat\beta_m/\sqrt{\Delta k}$ with $\Delta k=2\pi/N_B$. This way, in Hamiltonian \eqref{Ht} we can make the replacements
	\begin{align}
		\sum_{m=1}^{N_B} \omega^B_{m}\, \hat{\beta}^{\dagger}_{m}\hat{\beta}_{m}\rightarrow\int_0^\pi {\rm d}k\,\omega^B(k)\hat \beta^\dag(k)\hat \beta(k)\,,\,\,\,	\sum_{m'=1}^{N_B}\,\chi_{m'}\hat\beta_{m'} \rightarrow\,\int_0^\pi{\rm d} k\,\sqrt{\frac{2}{\pi}}\sin k \,\,\hat\beta(k)\,\,.\label{sinkk}
	\end{align}
Here, in each sum we multiplied and divided the summand by $\Delta k$, expressed it in terms of $\hat\beta_m/\sqrt{\Delta k}$ and finally carried out the continuous limit thus turning sums into integrals over the first Brillouin zone.

We assume now that the atom is tuned on resonance with a specific normal mode of block $A$ whose wavevector labeled by $m=m_0$, \ie 
\begin{equation}
\omega_0=\omega_c-2J \cos{k^A_{m_0}}\,. 
\end{equation}
Also, to ensure the weak-coupling regime, we assume that $g\ll J$ and that $\omega_0\equiv \omega_{m_0}$ is sufficiently far from the band edges $\omega_{c}\pm 2J$ (where singularities occur). Accordingly, we can effectively approximate the dispersion law of block $A$ to the first order around $m=m_0$ as
	\begin{align}
		\omega^A_m&\simeq\omega_0+v\left(k_{m}-k_{m_0}\right)=\omega_0 +v \frac{\pi}{N_A+1}\nu\,\,,\label{w0k0}
	\end{align}
where we used \eq\eqref{xim1} and replaced the effective group velocity $v= 2 J \sin k_{m_0}$. In the last identity, we introduced the integer number $\nu=m-m_0$ (taking both negative and positive values). Accordingly, we approximate $\xi_m$ [\cf\eq\eqref{xim1}] to the lowest order around $k_{m_0}$ obtaining
\begin{align}
		\xi_m \simeq J\sqrt{\frac{2}{N_A+1}}\,(-1)^{m}\sin{\left(k_{m_0}\right)}=\frac{v}{2}\sqrt{\frac{2}{N_A+1}}\,(-1)^{\nu}\,.
	\end{align}
Moreover, since the coupling strength between the generic block-$A$ mode and block B is much smaller than the waveguide bandwidth in the thermodynamic limit, \ie $\xi_m \ll J$ [\cf\eq\eqref{xim1}], we can also linearize the dispersion relation of block B as $\omega^B(k)\simeq\omega_0+v(k-k_0)$, where $v=2J \sin k_0$ and $k$ now runs between $-\infty$ and $+\infty$, which allows us to replace $\sin k$

	\subsection{Final continuous Hamiltonian}

	Putting everything together we get the total Hamiltonian
	\begin{align}
		\hat H_F&=\sum_{\nu=-\infty}^\infty\Omega_{\nu}\, \hat{\alpha}^{\dagger}_{\nu}\hat{\alpha}_{\nu}+\int_0^\pi {\rm d}k\,\omega(k)\hat \beta^\dag(k)\hat \beta(k)+\sum_{\nu=-\infty}^\infty \frac{v}{2}\sqrt{\frac{2}{N_A+1}}\,(-1)^{\nu}\,\left(\hat \alpha_\nu \int_0^\pi{\rm d} k\,\sqrt{\frac{2}{\pi}}\,\,\hat\beta^\dag(k)+{\rm H.c.}\right)\nonumber\\
		\hat V_{AF}&=\sum_{\nu=-\infty}^\infty g_\nu  \hat\sigma_{+}\hat\alpha_\nu+{\rm H.c.}\,,
	\end{align}
where $g_\nu $ is just \eqref{gm} expressed in terms of $\nu=m-m_0$ and $\phi=2 k_{m_0} n_0$.
	\begin{align}
	g_\nu&=g\, \sqrt{\frac{2}{N_A+1}}\sin{\left(\nu\frac{\pi}{N_A+1} n_0+ \frac{\phi}{2}\right)}\,\,,
\end{align}
	with $\Omega_\nu=\omega_0+ vk_\nu$ and $\omega(k)=\omega_0+ vk$.
	Each mode $\nu$ of the cavity is coupled to block-B with strength $\sim v/\sqrt{N_A}$, which for $N_A$ large enough will be far smaller than $v$. This and the fact that the coupling of each mode to block-B modes is flat (i.e., frequency-independent) yields that the integrals over $k$ can be extended to the entire real axis. By passing in addition to the frequency domain (using $\omega=v k$ and $\hat \beta(k)=\hat \beta (\omega)/\sqrt{v}$), we get the free-field Hamiltonian in the form
	\begin{align}
		\hat H_F&=\sum_{\nu} \Omega_{\nu}\, \hat{\alpha}^{\dagger}_{\nu}\hat{\alpha}_{\nu}+\int_{-\infty}^{\infty} {\rm d}\omega\,\omega\hat \beta^\dag(\omega)\hat \beta(\omega)+\sum_{\nu} \sqrt{\frac{v}{N_A+1}}\,\left(\hat \alpha_\nu \int_{-\infty}^\infty{\rm d} \omega\,\sqrt{\frac{1}{\pi}} \,\,\hat\beta^\dag(\omega)+{\rm H.c.}\right)\,,
	\end{align}
	where we also redefined the cavity modes as $\hat \alpha_{\nu}\rightarrow (-1)^\nu\hat \alpha_{\nu}$, in a way that factor $(-1)^{\nu}$ is now incorporated in the definition of the emitter-cavity.
	The latest step is to take the continuous limit of the tight binding model, which essentially leads to the replacement $N_A+1\rightarrow L_A$, which leads to \eqs\eqref{Heff}-\eqref{param2} in the main text.

\section{Spontaneous emission in the Markovian regime}
	
It is well-known that Hamiltonian \eqref{Hmodel} implies that the atom's excitation amplitude $\epsilon(t)$ obeys the exact delay differential equation \cite{DornerPRA02,TufarelliPRA13}
			\begin{equation}
			\dot \varepsilon(t)=-\frac{\Gamma}{2} \,\epsilon(t)+\frac{\Gamma}{2}e^{i\phi} \,\epsilon(t-\tau)\, \Theta (t-\tau)\,.\label{delay-de}
			\end{equation}
		{The exact solution of this equation is known and reads\cite{TufarelliPRA13}:
    \begin{equation}\label{eq_solutionspe}
         \varepsilon(t)=e^{-\frac{\Gamma}{2}t}\sum_n\frac{1}{n!}\left(\frac{\Gamma}{2}e^{i\phi+\frac{\Gamma}{2}\tau}\right)^n(t-n\tau)^n\theta(t-n\tau).  \end{equation}
       An approximate solution can be obtain}   for very short time delay (Markovian regime), by replacing $t-\tau$ $\simeq t$, so that \eqref{delay-de} reduces to
		    \begin{equation}
		    	\dot \epsilon=-\frac{\Gamma}{2} \,\epsilon+\frac{\Gamma}{2}e^{i\phi} \,\epsilon=i \frac{\Gamma}{2}\sin \phi\, \epsilon- \frac{\Gamma}{2}\left(1-\cos \phi\right) \epsilon\,.\label{delay-de2}
		    \end{equation}
	        Hence, the excited-state population will decay as $|\epsilon|^2=e^{-\Gamma' t}$ with 
         \begin{equation}
                      \Gamma'=\Gamma \left(1-\cos \phi\right)=2 \Gamma \sin^2 \frac{\phi}{2}\label{Gap}\,.
         \end{equation}

\subsection{One block-$A$ mode}

Consider the effective Hamiltonian \eqref{Heff} and approximate it by retaining only the block-$A$ resonant mode $\nu=0$, which is justified in the limit of very short time delay. Also, we take $L=x_0$, hence [\cf\eq\eqref{param2}] 
			\begin{equation}
			g_0=g\,\sqrt{\frac{2}{L}}\sin{\frac{\phi}{2}}\,\,.\label{g0}
			\end{equation}
			
			Let the atom initially in state $\ket{e}$ with mode $\nu=0$ and all the modes of block $B$ initially in the vacuum state. Then the joint state of the atom and mode $\nu=0$ has the form
			\begin{equation}
			\ket{\Psi(t)}=\epsilon (t)\ket{e,0}+a_0 (t)\ket{g,1}
		    \end{equation}
	        with $\ket{e,0}$ ($\ket{g,1}$) the state where the atom is in the excited (ground) state while mode $\nu=0$ has zero (one) photon, where $\epsilon (t)$ and $a_0 (t)$ fulfill the differential system (we set in a rotating frame such that $\omega_0=0$)
	        \begin{align}
	        	\dot\epsilon=-i g_0  \alpha_0\,\,\,\,\,	\dot\alpha_0=- \tfrac{\gamma}{2} \alpha_0-i g_0 \epsilon\nonumber
	        \end{align}
        	subject to the intial condition $\epsilon(0)=1$, $\alpha_0(0)=0$.
           \\
           \\
           In the Laplace domain (variable $t$ replaced by $s$), the system reads
           \begin{align}
           	s\tilde\epsilon-1=-i g_0  \tilde\alpha_0\,\,\,\,\, s\tilde\alpha_0=- \tfrac{\gamma}{2} \tilde\alpha_0-i g_0 \tilde\epsilon\label{epsa}\,.
           \end{align}
In the limit of very short delay, we get $\gamma\gg g_0$ [\cf\eqs\eqref{param1}-\eqref{param2}]; hence we can replace $\alpha(t)$ with its stationary value for given $\epsilon(t)$ which is equivalent to setting $s=0$ in the second identity of \eq\eqref{epsa}. This yields $\tilde\alpha_0=-i \frac{2g_0}{\gamma}\tilde\epsilon$. Replacing in the equation for $\tilde{\epsilon}$ we get
       \begin{equation}
       	\tilde\epsilon=\frac{1}{s+\Gamma'}\,,\label{eps2}
       \end{equation}
where $\Gamma'=\frac{2 g_0^2}{\gamma}$ matches \eq\eqref{Gap}.

{\subsection{Number of cavity modes needed to capture the non-Markovian dynamics in \fig\ref{fig-test}(b)}}

{To assess how many modes $N_A$ of block $A$ (fictitious cavity) must be retained in order to accurately capture the non-Markovian behavior, we focus on the paradigmatic process of spontaneous emission in which case the exact solution for the atomic excited-state amplitude versus time is given in Eq.~\eqref{eq_solutionspe}. This is to be compared with the approximate solution obtained from the effective model for different values of the parameters $\Gamma \tau$ and $\phi$.
}
{To carry out this task, for each set of parameters we discretize time as $t_m=m \Delta t$, with $\Delta t$ the time step, up to a final time $t_{\rm f}=N_T\Delta t$ with $N_T$ the total number of time steps. The exact and approximate solutions are accordingly represented by the $N_T$-dimensional vectors of components $\varepsilon(t_m)$ and $\varepsilon_{\rm eff}(t_m)$, respectively.}

{As a figure of merit quantifying the mismatch between the exact and approximates solutions, we use the root mean square deviation  the standard deviation defined by
\begin{equation}
    \mathrm{RMSD} = \sqrt{\frac{1}{N_t} \sum_{m=1}^{N_t} \left[ \varepsilon_{\mathrm{eff}}(t_m) - \varepsilon(t_m) \right]^2}\,.
\end{equation}
We then run simulations using the effective model for growing values of $N_A$ and compute the RMSD for each value of $N_A$. The minimum value of $N_A$ yielding a root mean square deviation below $1\%$, i.e., such that $\mathrm{RMSD} < 0.01$, then embodies our measure for the required number of cavity modes.}
\\
\\
\bigskip

\section{Tensor Network simulations}

\begin{figure}
	\includegraphics[width=0.8 \textwidth]{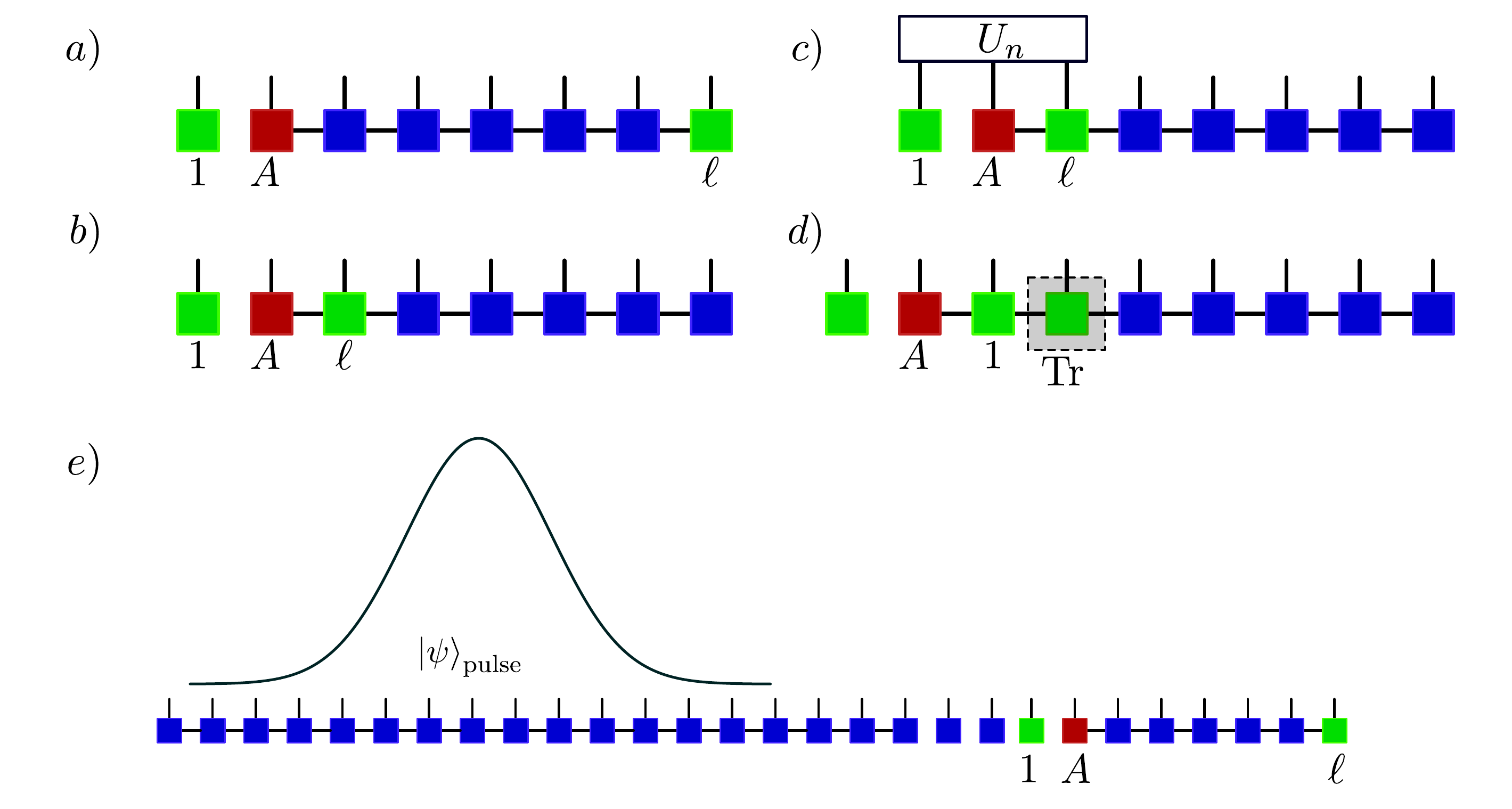}
	\caption{Update of the MPO representing the joint state of the atom and the non-Markovian bath according to \eqref{s_eq:mpo}. The tensor in red represents $A^{i_A,i'_A}$, indicating the atom's state. The blue tensors correspond to the chain's harmonic oscillators, while the green ones represent the first and $\ell$th oscillators of the chain, which interact with the atom. Each physical leg denotes the pair of vector spaces corresponding to each tensor. Consequently, the operators acting on the chain should be understood as the corresponding maps, or equivalently, the operators acting on the local purification form matrix product state (PMPS) of the system's density matrix \cite{Cuevas_2013}. a) After the $n-1$th step the joint atom-bath system and the oscillator $1$ on the left are uncorrelated. b) The $\ell$th oscillator is put on the right of the atom through a sequence of swap operations (not depicted). c) the tensor network corresponding to the propagator \eqref{s_eq:prop} acts on the systems $1, A$ and $\ell$. d) After the interaction, the $\ell$th oscillator is traced, $A$ and 1 swap their positions, and a new uncorrelated oscillator is added to the left. A compression follows after each transformation of the chain.
    e) Initial state for the scattering dynamics. A chain of oscillators encoding the state of the incoming pulse is placed on the left of the joint atom-bath system. 
    }\label{fig-mps}
\end{figure}

The exact results throughout the main text are obtained by leveraging the MPS formalism specifically adapted for photonic circuits featuring time delays, as illustrated in \cite{pichlerPhotonic2016}.
We consider the reduced quantum state of the emitter $A$ and the non-Markovian bath $F$ (field in the region within the atom and the mirror) as initially uncorrelated, i.e. $\rho_{AF}(t=0) = \rho_A (0) \otimes \rho_F (0)$. For a single emitter coupled to a semi-infinite waveguide, a simplification of the problem's geometry arises by transforming the configuration into an equivalent one with a chiral infinite waveguide and the emitter coupled to the waveguide at two points separated by a distance of $2 x_0$ \cite{FangNJP18}.
In the interaction picture with respect to the bath and the emitter free Hamiltonians,  \eq\eqref{Hmodel} is can be recast as \cite{DarioTB}:
\begin{equation}
\hat H (t) = \hat{H}_{\rm dr} + g  (\,\hat \sigma^{+}_A ( {\hat b}_{t} + e^{i 2 \omega_0 \tau} \, {\hat b}_{t-2\tau} )+ {\rm H.c.} )\,\label{Vt1}\,
\end{equation}
where we are including a classical driving Hamiltonian on the atom $\hat{H}_{\rm dr} = \Omega (\sigma^{+}_A + \sigma^{-}_A)$ and ${\hat b}_{t}$ are the time-domain ladder operator of the chiral bath.
The overall state $\rho_{AF}$ is evolved according to a stroboscopic map with discrete time steps $\Delta t$ chosen to be small compared to the relevant frequencies of the system, consistent with the approach used in quantum collision models \cite{Ciccarello_2022}.
Hence we define the discrete-time propagator
\begin{align}
\label{s_eq:prop}
\hat{U}_n = \exp{-i \hat{H}_n \Delta t}\,,
\end{align}
with 
\begin{align}
\hat{H}_n = \hat{H}_{\rm dr} + \frac{g}{\sqrt{\Delta t}}  (\,\hat \sigma^{+}_A ( {\hat b}_{n} + e^{i \phi} \, {\hat b}_{n-\ell} )+ {\rm H.c.} ) \,,
\end{align}
where we introduced the discrete bosonic noise operators $\hat{b}_n = \frac{1}{\sqrt{\Delta t}} \int_{t_{n-1}}^{t_{n}} \hat{b}(s) ds$ and $\ell=2 \tau/{\Delta t}$.
The discretization of the interaction reflects in the representation of the environment as a chain of $\ell$ quantum harmonic oscillators and the joint state of atom and environment as
\begin{align}
\rho_{AF} = \sum_{\overline{i},\overline{i}'} c_{\overline{i},\overline{i}'} |\overline{i} \rangle \langle \overline{i}' |
\end{align}
with the basis state $|\overline{i}\rangle = | i_A, i_{1}, \ldots, i_{{\ell}} \rangle$, where the numbers identify the $\ell$ oscillators.
This representation can be reformulated using singular value decomposition between each possible bipartition of the chain, yielding a Matrix Product Operator (MPO)
\begin{align}
\rho_{AF} = \sum_{\overline{i},\overline{i}'} \sum_{\overline{\kappa}} A^{i_A,i'_A}_{\kappa_1}
A^{i_{1},i'_{1}}_{\kappa_1,\kappa_2}\ldots A^{i_{\ell},i'_{\ell}}_{\kappa_{\ell-1}} |\overline{i} \rangle \langle \overline{i}' |\,.
\label{s_eq:mpo}
\end{align}
Here, the indices $\overline{i}$ and $\overline{i}'$ iterate over the computational basis of each subsystem (\textit{physical indices}), and the contracted indices $\overline{\kappa}$ (\textit{virtual} or \textit{bond indices}) run from $0$ to a maximum value $D_{\rm max}$ called \textit{bond dimension}, capturing correlations between the sites \cite{SchollwockAnnPhys11,VidalPRL03}.
Truncating the bond dimension up to a certain threshold corresponds to discarding the smallest singular values in the decomposition mentioned above. This approximation proves highly effective, especially when the subsystems exhibit weak correlations, significantly reducing the computational resources required to manage the state. Any operator can be expressed as a Matrix Product Operator (MPO) and operates on the state by contracting the relevant physical indices. 
After the action of a non-local operator, the bond dimension between two sites involved in the evolution typically increases. Therefore, a \textit{compression} step, i.e., singular value decomposition followed by truncation of the bond dimension, is always performed to effectively manage the dimension of the tensor network.
In particular, the propagator \eqref{s_eq:prop} features terms acting on $A$ and on two oscillators (the $1$st and the $\ell$th) at the same time. Given the high cost of compression for long-range interactions, we mitigate this by simplifying the propagator to the application of nearest neighbor unitary operations. This is achieved through a suitable swap scheme that maintains the physics unaltered \cite{pichlerPhotonic2016,Schachenmayer_2010}. The complete process of evolution-update is illustrated in \fig\ref{fig-mps} a-d, utilizing the Penrose notation for tensors.
This scheme can be directly applied to spontaneous-emission dynamics without modification.
For scattering dynamics, an additional step is necessary to represent the incident pulse because the incoming field constitutes a correlated state across multiple oscillators.
Let $m$ be the number of such oscillators. Thus we define the discrete n-particle wavepacket operator as
\begin{align}
\hat{\psi}_n^\dag = (\sum_{i=0}^m \sqrt{\Delta t} \,\xi_i \, \hat{a}_i^\dag)^n \,,
\label{s_eq:n_ph_op}
\end{align}
where $\xi_i$ is a discrete sample of the Gaussian amplitude
\begin{align}\label{eq:gaussian}
\xi(t)_{t_0} = \left(\frac{W^2}{2 \pi}\right)^{1/4} \exp \left\{\frac{1}{4} W^2 (t-t_0)^2 \right\}\,,
\end{align}
with $W$ and $t_0$ the frequency bandwidth and the center of the pulse respectively.
The coherent-state pulse with average photon number $|\alpha|^2$ reads \cite{Zheng2010}
\begin{align}
\ket{\psi}_{\rm pulse} = e^{-|\alpha|^2/2}\sum_{n=0}^{\infty} \frac{\alpha^n}{n!} \hat{\psi}_n^\dag \ket{0}\,.
\end{align}
This state can be generated as a Matrix Product Operator (MPO) by applying the tensor network corresponding to \eqref{s_eq:n_ph_op} to a chain of $m$ oscillators initially in the vacuum state. For low powers, the first terms of the sum above are sufficient to represent the pulse. 
The atomic population, $\varepsilon$,  the output field intensity$I_{\rm out}$, and the two-photon correlation function,
 $G^{(2)}$,  are obtained through measurements on the reduced state of the atom, $\hat\rho_A (t) = {\rm Tr}_{\rm F}\{ \hat\rho_{\rm AF}(t)\}$, and of the $\ell$th oscillator, $\hat\rho_\ell (t) = {\rm Tr}_{\rm AF \backslash \ell}\{ \hat\rho_{\rm AF}(t)\}$ respectively:
\begin{align}
\varepsilon (t) &= {\rm Tr}\{ \hat\rho_A (t) \hat\s_+ \hat\s_- \} \,,\\
I_{\rm out} (t) &= {\rm Tr}\{ \hat\rho_\ell (t) \hat{a}^{\dagger}_{\ell}(t) \hat{a}_{\ell}(t) \} \,,\\
G^{(2)} (t) & = {\rm Tr}\{ \hat\rho_\ell (t) \hat{a}^{\dagger}_{\ell}(t) \hat{a}^{\dagger}_{\ell}(t) \hat{a}_{\ell}(t) \hat{a}_{\ell}(t) \} \,.
\end{align}

\section{Input output formalism for the effective model}
To simulate the scattering of a photonic wavepacket with the emitter within our effective model we treat the block B waveguide modes, $\hat\beta_\omega$, as a Markovian bath. In this way is possible to eliminate these degrees of freedom obtaining the following master equation for an open multimode cavity QED system
\begin{equation}\label{eq:Me_scattering}
	\dot{\hat\rho} = -i[H_{\rm CQED}+H_D, \hat\rho]
	+\sqrt{\gamma} \mathcal{D}[\hat A]\hat\rho \,,
\end{equation}
where 
\begin{equation}
H_{\rm CQED}=\omega_0\, \hat\sigma_+\hat\sigma_-{+}\sum_\nu \Omega_{\nu}\, {\hat\alpha}^{\dagger}_{\nu}{\hat\alpha}_{\nu}+\sum_\nu g_\nu \,( \hat\alpha_\nu^\dag \hat\sigma_-+{\rm H.c.})  
\end{equation}
is the system Hamiltonian with $\Omega_{\nu}$ and $g_\nu$ being defined in the main text. In Eq. \eqref{eq:Me_scattering} we introduced the usual Linbladian dissipator 
$\mathcal{D}[\hat A]\rho=\hat A\rho\hat A^\dag-\{\hat A^\dag \hat A,\rho\}/2$
 applied to the collective mode operator $\hat A=\sum_\nu\hat\alpha_{\nu}$ with $\gamma=2 v/L$ being the decay rate of block A into the block B bath. In Eq. \eqref{eq:Me_scattering} we included   a coherent  driving term of the modes $\alpha_\nu$: 
\begin{equation}
H_D=\sqrt{\gamma}\sum_{\nu}[E_{\rm in}(t)e^{i\omega_{\rm in} t}\hat\alpha^\dagger_\nu +\rm H.c]
\end{equation}
where $\omega_{\rm in}$  is  the frequency   of the driving  input field.  
The shape of the input pulse is determined by the field amplitude $E_{\rm in}(t)$, normalized with respect to the number of photons $n_{\rm ph}$, $\int dt |E_{\rm in}(t)|^2=n_{\rm ph}$. For the scattering process discussed in the main text, we employ the same Gaussian pulse shape as specified in Eq. \eqref{eq:gaussian},  $E_{\rm in}(t):=\xi(t)_{t_0}$.
The open-system dynamics of the multimode  cavity QED system, described by Eq. \eqref{eq:Me_scattering}, is simulated using a quantum trajectories approach, averaging over $n_t=4000$ trajectories \cite{molmer1993monte}.
Once solved the system dynamics 
the output field can be reconstructed using the following input-output equation \cite{gardiner1985input}:
\begin{equation}
      \hat E_{\rm out}(t) =\hat E_{\rm in}(t)+i\sqrt{\gamma}\sum_\nu\hat\alpha_{\nu}(t).
  \end{equation}
The intensity of the output field and the two-photon correlation function can then be computed in terms of the modes of the effective cavity and read $I_{\rm out}(t)=\langle \hat\alpha^{\dagger}_{\nu}(t)\hat\alpha_{\nu}(t)\rangle$ and $G^{(2)}(t)=\langle \hat\alpha^{\dagger}_{\nu}(t)\hat\alpha^{\dagger}_{\nu}(t)\hat\alpha_{\nu}(t)\hat\alpha_{\nu}(t)\rangle$, respectively.

{
\section{Multi-atom generalization}
\label{app:E}
The effective model introduced in the main text for one atom and a terminated waveguide can be generalized to the case of many atoms emitting into a waveguide. For the sake of argument, we consider two atoms in an infinite waveguide; the extension to a terminated waveguide and/or many atoms is straightforward.
}
The method is based on decomposing the waveguide into a modular arrangement of fictitious cavities, one for each emitter. 

\subsection*{Review of the one-atom case for a semi-infinite waveguide}

The single-atom configuration in the main text features consider an atom located at a distance $x_0$ from a perfect mirror terminating a semi-infinite waveguide [see \fig\ref{fig-setup}(a)]. The effective model is obtained by isolating a segment of the waveguide of length $L > x_0$, which defines a fictitious cavity bounded by the mirror on left side and coupled to an external Markovian reservoir representing the rest of the waveguide. The cavity supports a discrete set of modes with frequencies defined in Eq \eqref{param1}.
The atom couples to these modes with frequency-dependent strengths \eqref{param2}.
We observe that convergence of the effective model to the exact atom–waveguide dynamics depends on the choice of $\alpha = L / x_0$. In this setting, the best results are obtained when $\alpha = 2$, i.e., when the atom is placed at the center of the fictitious cavity. This can be understood by noting that the amplitude of all odd-parity cavity modes have anti-nodes at the cavity center, leading to constructive interference and maximum overlap with the atomic position. In contrast, arbitrary values of $\alpha$ generally lead to partial cancellation of modal contributions.

\subsection*{Two atoms in an infinite waveguide}

We consider two atoms separated by a distance $x_0$ coupled to an infinite waveguide described by the Hamiltonian
\begin{align}\label{Hmodel-2}
	H&=\omega_0\, \sum_{i=1,2}\hat\sigma_{i+}\hat\sigma_{i-} \!-iv\!\!\int_{-\infty}^{\infty}\!\! \!{\rm d} x\!\left[ \hat a^{\dagger}_R(x)\partial_x \hat a_R(x)\!-\! \hat a^{\dagger}_L(x)\partial_x \hat a_L(x)\right] + g\! \sum_{i=1,2}\int_{-\infty}^{\infty} \!\!\!\!{\rm d} x \left[ \hat\sigma_+\left( \hat a_{ L}(x)+ \hat a_{ R}(x)\right)\!+\!{\rm H.c.}\right]\delta(x{-}x_i)\,,\!\!\!
\end{align}
with $x_1=-x_2=-x_0/2$ defining the atom positions (we set the origin at the midpoint between the two emitters). The relevant parameters of the system are now [\cf\eq\eqref{param0}] $\Gamma={2g^2}/{v}$, $\tau= {2 x_0}/{v}$ and $\phi=2k_0x_0$.

\begin{figure}
	\includegraphics[width=0.7 \textwidth]{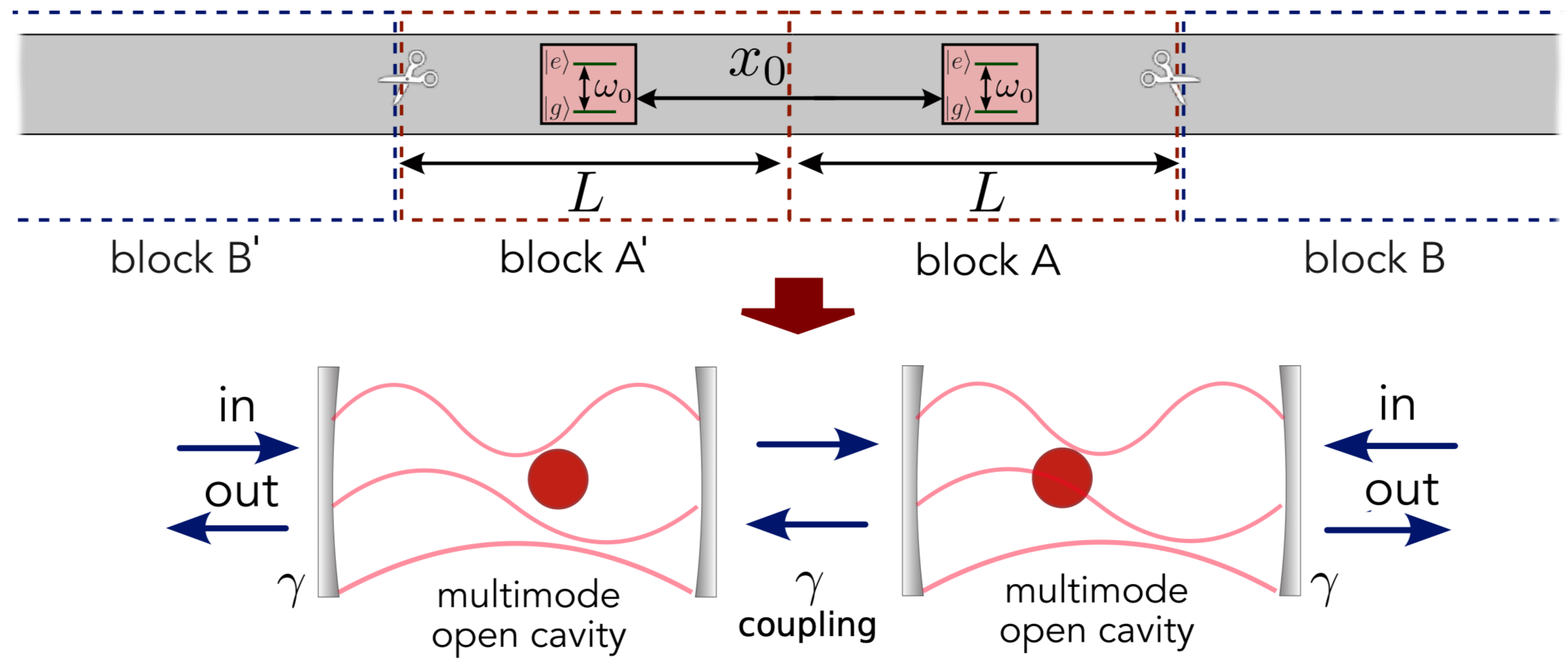}
	\caption{Generalization of the framework for two atoms in an infinite waveguide.  
The waveguide is decomposed into four blocks. Each of the blocks \( A \) and \( A' \) is treated as a multi-mode open cavity of length \( L \) coupled to one of the atoms. Blocks $A$ and $A'$ additionally leak into blocks \( B \) and \( B' \), respectively. Each pair of adjacent blocks is coupled with rate \( \gamma \).  
}

\label{fig-setup2atoms}
\end{figure}

To extend the effective model of the main text to the present case, we somewhat replicate the one-atom scheme by introducing two fictitious cavities $A$ and $A'$, each containing one atom and treated as a multi-mode open cavity of length \( L \) (see Fig.~\ref{fig-setup2atoms}). The two cavities are connected through a fictitious mirror, analogous to the one coupling blocks $A$ and $B$ in the single-atom configuration of \fig\ref{fig-setup}(a).
Blocks $A$ and $A'$ respectively leak into blocks \( B \) and \( B' \) embodying Markovian baths. Each pair of adjacent blocks is coupled with rate \( \gamma \).
The effective Hamiltonian corresponding to such block decomposition thus reads

\begin{align}
		H_{\rm eff}=&\,\, \omega_0\, \sum_{i=1,2}\hat\sigma_{i+}\hat\sigma_{i-} {+}\sum_\nu \Omega_{\nu}\, ({\hat\alpha}^{\dagger}_{\nu}{\hat\alpha}_{\nu} + {\hat\alpha}'^{\dagger}_{\nu}{\hat\alpha}'_{\nu} ){+}\int\!{\rm d}\omega\,\omega\, (\hat\beta^\dag_\omega \hat\beta_\omega + \hat\beta'^\dag_\omega \hat\beta'_\omega ) + \sqrt{\frac{\gamma}{2\pi}}\,\sum_\nu\int\! {\rm d} \omega \,(  \hat\alpha_\nu^\dag  \hat\beta_\omega +  \hat\alpha_\nu'^\dag  \hat\beta'_\omega  +  {\rm H.c.}) \\& 
        +  \sum_\nu g_\nu \,( \hat\alpha_\nu^\dag \hat\sigma_{1-} + \hat\alpha_\nu'^\dag \hat\sigma_{2-} + {\rm H.c.})
        + \sqrt{\frac{\gamma}{2\pi}}\,\sum_{\nu \nu'} (\hat\alpha_\nu^\dag  \hat\alpha'_{\nu'} +  {\rm H.c.} )
        \notag
        \label{Heff}\,,
	\end{align}
where $\hat{\alpha}_\nu$ ($\hat{\alpha}'_\nu$) are normal-mode ladder operators of block $A$ ($A'$) and likewise $\hat{\beta}_\omega$ ( $\hat{\beta}'_\omega$) correspond to normal modes of block $B$ ($B'$). The last term represents the coupling between blocks $A$ and $A'$. The atom position in the corresponding block is now defined by the dimensionless parameter $\alpha=L/(x_0/2)$.

\begin{figure}
\label{fig-2at}
	\includegraphics[width=0.75 \textwidth]{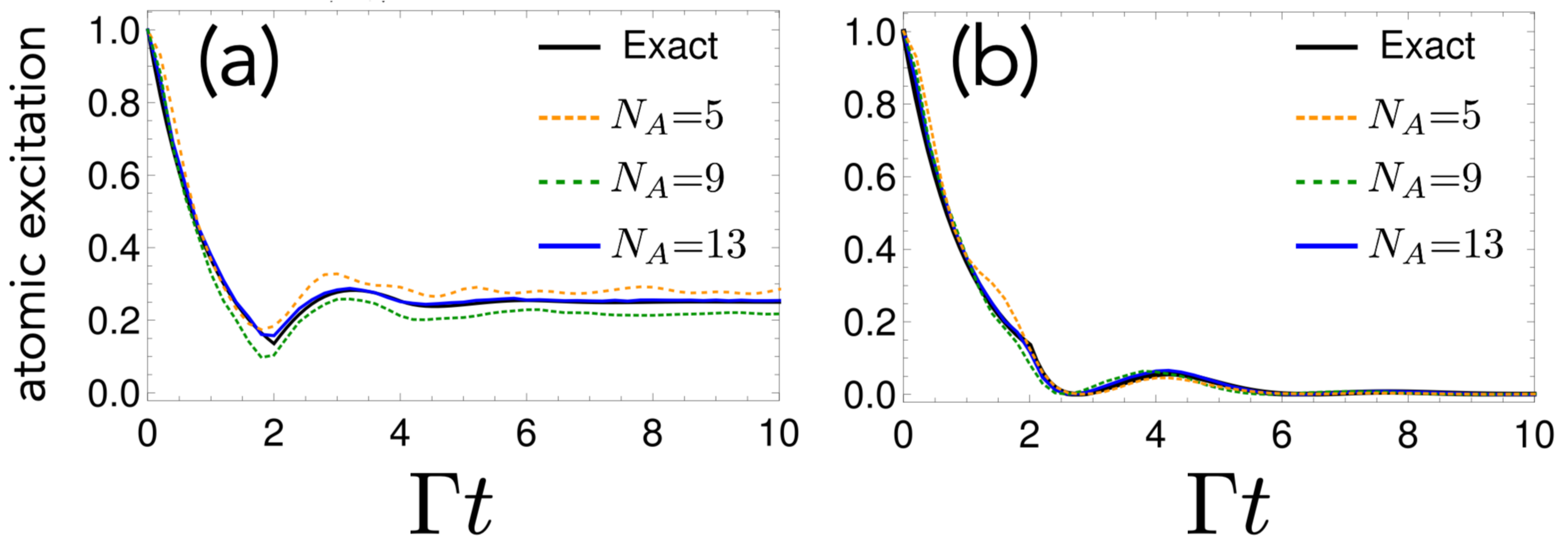}
	\caption{Total atomic excitation versus time for two atoms initially in the sub-radiant state $\ket{\psi_{\rm sub}}=\tfrac{1}{\sqrt{2}}\left(\ket{eg}_{12}{-}\ket{ge}_{12}\right)$
     (a) and in the super-radiant state $\ket{\psi_{\rm sup}}=\tfrac{1}{\sqrt{2}}\left(\ket{eg}_{12}{+}\ket{ge}_{12}\right)$
     (b) for different numbers of block modes $N_A=N_{A'}$. We set $\phi=\pi$, $\Gamma\tau=2$ and $\alpha=3/2$.}    
\end{figure}

In this modular framework, the total propagation phase $\phi$ is effectively distributed between the two cavities. To maintain exact correspondence with the full waveguide dynamics, we incorporated an additional $\pi/2$-phase shift in the definition of the atom–cavity coupling function for each block. This ensures that the effective model accurately reproduces the correct dynamics for any value of $\phi$.
For a single atom in a semi-infinite waveguide, placing the emitter at the center of a hard-wall cavity aligns it with the antinodes of all odd modes, maximizing the coupling. Simply joining two such cavities, however, imposes a node at the midpoint between the atoms, which does not capture the full dynamics. To fix this, we instead select modes with an antinode at the interface, introducing a $\pi/2$ phase shift in $g$. This choice, corresponding to $\alpha = 3/2$, reproduces the correct geometry of the coupling points for the two atoms. 
\bigskip

To test the effectiveness of the present multi-atom framework, in \fig\ref{fig-2at} we consider the spontaneous emission of two emitters prepared in the super-(sub-) radiant state $\ket{\psi_{\rm sup/sub}}=\tfrac{1}{\sqrt{2}}\left(\ket{eg}_{12}{\pm}\ket{ge}_{12}\right)$ for $\phi=\pi$, $\alpha=3/2$, the relatively long delay $\Gamma\tau=2$ and for growing values of $N_A=N_{A'}$. Similarly to \fig\ref{fig-test}(a), the approximate solution converges to the exact one \cite{SinhaPRL20} for a sufficiently large number of modes.
\\

The present modular configuration is naturally generalized to a larger numbers of atoms by defining as many fictitious adjacent blocks, each containing one atom.

\putbib
\end{bibunit}


\begin{thebibliography}{0}%
\makeatletter
\providecommand \@ifxundefined [1]{%
 \@ifx{#1\undefined}
}%
\providecommand \@ifnum [1]{%
 \ifnum #1\expandafter \@firstoftwo
 \else \expandafter \@secondoftwo
 \fi
}%
\providecommand \@ifx [1]{%
 \ifx #1\expandafter \@firstoftwo
 \else \expandafter \@secondoftwo
 \fi
}%
\providecommand \natexlab [1]{#1}%
\providecommand \enquote  [1]{``#1''}%
\providecommand \bibnamefont  [1]{#1}%
\providecommand \bibfnamefont [1]{#1}%
\providecommand \citenamefont [1]{#1}%
\providecommand \href@noop [0]{\@secondoftwo}%
\providecommand \href [0]{\begingroup \@sanitize@url \@href}%
\providecommand \@href[1]{\@@startlink{#1}\@@href}%
\providecommand \@@href[1]{\endgroup#1\@@endlink}%
\providecommand \@sanitize@url [0]{\catcode `\\12\catcode `\$12\catcode `\&12\catcode `\#12\catcode `\^12\catcode `\_12\catcode `\%12\relax}%
\providecommand \@@startlink[1]{}%
\providecommand \@@endlink[0]{}%
\providecommand \url  [0]{\begingroup\@sanitize@url \@url }%
\providecommand \@url [1]{\endgroup\@href {#1}{\urlprefix }}%
\providecommand \urlprefix  [0]{URL }%
\providecommand \Eprint [0]{\href }%
\providecommand \doibase [0]{https://doi.org/}%
\providecommand \selectlanguage [0]{\@gobble}%
\providecommand \bibinfo  [0]{\@secondoftwo}%
\providecommand \bibfield  [0]{\@secondoftwo}%
\providecommand \translation [1]{[#1]}%
\providecommand \BibitemOpen [0]{}%
\providecommand \bibitemStop [0]{}%
\providecommand \bibitemNoStop [0]{.\EOS\space}%
\providecommand \EOS [0]{\spacefactor3000\relax}%
\providecommand \BibitemShut  [1]{\csname bibitem#1\endcsname}%
\let\auto@bib@innerbib\@empty
\end{thebibliography}%


\begin{thebibliography}{69}%
\makeatletter
\providecommand \@ifxundefined [1]{%
 \@ifx{#1\undefined}
}%
\providecommand \@ifnum [1]{%
 \ifnum #1\expandafter \@firstoftwo
 \else \expandafter \@secondoftwo
 \fi
}%
\providecommand \@ifx [1]{%
 \ifx #1\expandafter \@firstoftwo
 \else \expandafter \@secondoftwo
 \fi
}%
\providecommand \natexlab [1]{#1}%
\providecommand \enquote  [1]{``#1''}%
\providecommand \bibnamefont  [1]{#1}%
\providecommand \bibfnamefont [1]{#1}%
\providecommand \citenamefont [1]{#1}%
\providecommand \href@noop [0]{\@secondoftwo}%
\providecommand \href [0]{\begingroup \@sanitize@url \@href}%
\providecommand \@href[1]{\@@startlink{#1}\@@href}%
\providecommand \@@href[1]{\endgroup#1\@@endlink}%
\providecommand \@sanitize@url [0]{\catcode `\\12\catcode `\$12\catcode `\&12\catcode `\#12\catcode `\^12\catcode `\_12\catcode `\%12\relax}%
\providecommand \@@startlink[1]{}%
\providecommand \@@endlink[0]{}%
\providecommand \url  [0]{\begingroup\@sanitize@url \@url }%
\providecommand \@url [1]{\endgroup\@href {#1}{\urlprefix }}%
\providecommand \urlprefix  [0]{URL }%
\providecommand \Eprint [0]{\href }%
\providecommand \doibase [0]{https://doi.org/}%
\providecommand \selectlanguage [0]{\@gobble}%
\providecommand \bibinfo  [0]{\@secondoftwo}%
\providecommand \bibfield  [0]{\@secondoftwo}%
\providecommand \translation [1]{[#1]}%
\providecommand \BibitemOpen [0]{}%
\providecommand \bibitemStop [0]{}%
\providecommand \bibitemNoStop [0]{.\EOS\space}%
\providecommand \EOS [0]{\spacefactor3000\relax}%
\providecommand \BibitemShut  [1]{\csname bibitem#1\endcsname}%
\let\auto@bib@innerbib\@empty
\bibitem [{\citenamefont {Gardiner}\ and\ \citenamefont {Zoller}(2004)}]{gardiner2004}%
  \BibitemOpen
  \bibfield  {author} {\bibinfo {author} {\bibfnamefont {C.}~\bibnamefont {Gardiner}}\ and\ \bibinfo {author} {\bibfnamefont {P.}~\bibnamefont {Zoller}},\ }\href {https://www.springer.com/gp/book/9783540223016} {\emph {\bibinfo {title} {Quantum {Noise}: {A} {Handbook} of {Markovian} and {Non}-{Markovian} {Quantum} {Stochastic} {Methods} with {Applications} to {Quantum} {Optics}}}},\ \bibinfo {edition} {3rd}\ ed.,\ Springer {Series} in {Synergetics}\ (\bibinfo  {publisher} {Springer-Verlag},\ \bibinfo {address} {Berlin Heidelberg},\ \bibinfo {year} {2004})\BibitemShut {NoStop}%
\bibitem [{\citenamefont {Liao}\ \emph {et~al.}(2016)\citenamefont {Liao}, \citenamefont {Zeng}, \citenamefont {Nha},\ and\ \citenamefont {Zubairy}}]{LiaoPhyScr16}%
  \BibitemOpen
  \bibfield  {author} {\bibinfo {author} {\bibfnamefont {Z.}~\bibnamefont {Liao}}, \bibinfo {author} {\bibfnamefont {X.}~\bibnamefont {Zeng}}, \bibinfo {author} {\bibfnamefont {H.}~\bibnamefont {Nha}},\ and\ \bibinfo {author} {\bibfnamefont {M.~S.}\ \bibnamefont {Zubairy}},\ }\bibfield  {title} {\emph {\bibinfo {title} {{Photon transport in a one-dimensional nanophotonic waveguide QED system}}},\ }\href {https://doi.org/10.1088/0031-8949/91/6/063004} {\bibfield  {journal} {\bibinfo  {journal} {Physica Scripta}\ }\textbf {\bibinfo {volume} {91}},\ \bibinfo {pages} {63004} (\bibinfo {year} {2016})}\BibitemShut {NoStop}%
\bibitem [{\citenamefont {Roy}\ \emph {et~al.}(2017)\citenamefont {Roy}, \citenamefont {Wilson},\ and\ \citenamefont {Firstenberg}}]{RoyRMP17}%
  \BibitemOpen
  \bibfield  {author} {\bibinfo {author} {\bibfnamefont {D.}~\bibnamefont {Roy}}, \bibinfo {author} {\bibfnamefont {C.~M.}\ \bibnamefont {Wilson}},\ and\ \bibinfo {author} {\bibfnamefont {O.}~\bibnamefont {Firstenberg}},\ }\bibfield  {title} {\emph {\bibinfo {title} {{Colloquium: Strongly interacting photons in one-dimensional continuum}}},\ }\href {https://doi.org/10.1103/RevModPhys.89.021001} {\bibfield  {journal} {\bibinfo  {journal} {Reviews of Modern Physics}\ }\textbf {\bibinfo {volume} {89}},\ \bibinfo {pages} {21001} (\bibinfo {year} {2017})}\BibitemShut {NoStop}%
\bibitem [{\citenamefont {Sheremet}\ \emph {et~al.}(2023)\citenamefont {Sheremet}, \citenamefont {Petrov}, \citenamefont {Iorsh}, \citenamefont {Poshakinskiy},\ and\ \citenamefont {Poddubny}}]{Sheremet-RMP}%
  \BibitemOpen
  \bibfield  {author} {\bibinfo {author} {\bibfnamefont {A.~S.}\ \bibnamefont {Sheremet}}, \bibinfo {author} {\bibfnamefont {M.~I.}\ \bibnamefont {Petrov}}, \bibinfo {author} {\bibfnamefont {I.~V.}\ \bibnamefont {Iorsh}}, \bibinfo {author} {\bibfnamefont {A.~V.}\ \bibnamefont {Poshakinskiy}},\ and\ \bibinfo {author} {\bibfnamefont {A.~N.}\ \bibnamefont {Poddubny}},\ }\bibfield  {title} {\emph {\bibinfo {title} {Waveguide quantum electrodynamics: Collective radiance and photon-photon correlations}},\ }\href {https://doi.org/10.1103/RevModPhys.95.015002} {\bibfield  {journal} {\bibinfo  {journal} {Rev. Mod. Phys.}\ }\textbf {\bibinfo {volume} {95}},\ \bibinfo {pages} {015002} (\bibinfo {year} {2023})}\BibitemShut {NoStop}%
\bibitem [{\citenamefont {Andersson}\ \emph {et~al.}(2019)\citenamefont {Andersson}, \citenamefont {Suri}, \citenamefont {Guo}, \citenamefont {Aref},\ and\ \citenamefont {Delsing}}]{andersson2019non}%
  \BibitemOpen
  \bibfield  {author} {\bibinfo {author} {\bibfnamefont {G.}~\bibnamefont {Andersson}}, \bibinfo {author} {\bibfnamefont {B.}~\bibnamefont {Suri}}, \bibinfo {author} {\bibfnamefont {L.}~\bibnamefont {Guo}}, \bibinfo {author} {\bibfnamefont {T.}~\bibnamefont {Aref}},\ and\ \bibinfo {author} {\bibfnamefont {P.}~\bibnamefont {Delsing}},\ }\bibfield  {title} {\emph {\bibinfo {title} {Non-exponential decay of a giant artificial atom}},\ }\href {https://doi.org/10.1038/s41567-019-0605-6} {\bibfield  {journal} {\bibinfo  {journal} {Nature Physics}\ }\textbf {\bibinfo {volume} {15}},\ \bibinfo {pages} {1123–1127} (\bibinfo {year} {2019})}\BibitemShut {NoStop}%
\bibitem [{\citenamefont {Ferreira}\ \emph {et~al.}(2021)\citenamefont {Ferreira}, \citenamefont {Banker}, \citenamefont {Sipahigil}, \citenamefont {Matheny}, \citenamefont {Keller}, \citenamefont {Kim}, \citenamefont {Mirhosseini},\ and\ \citenamefont {Painter}}]{PainterPRX21}%
  \BibitemOpen
  \bibfield  {author} {\bibinfo {author} {\bibfnamefont {V.~S.}\ \bibnamefont {Ferreira}}, \bibinfo {author} {\bibfnamefont {J.}~\bibnamefont {Banker}}, \bibinfo {author} {\bibfnamefont {A.}~\bibnamefont {Sipahigil}}, \bibinfo {author} {\bibfnamefont {M.~H.}\ \bibnamefont {Matheny}}, \bibinfo {author} {\bibfnamefont {A.~J.}\ \bibnamefont {Keller}}, \bibinfo {author} {\bibfnamefont {E.}~\bibnamefont {Kim}}, \bibinfo {author} {\bibfnamefont {M.}~\bibnamefont {Mirhosseini}},\ and\ \bibinfo {author} {\bibfnamefont {O.}~\bibnamefont {Painter}},\ }\bibfield  {title} {\emph {\bibinfo {title} {Collapse and revival of an artificial atom coupled to a structured photonic reservoir}},\ }\href {https://doi.org/10.1103/PhysRevX.11.041043} {\bibfield  {journal} {\bibinfo  {journal} {Phys. Rev. X}\ }\textbf {\bibinfo {volume} {11}},\ \bibinfo {pages} {041043} (\bibinfo {year} {2021})}\BibitemShut {NoStop}%
\bibitem [{\citenamefont {Lechner}\ \emph {et~al.}(2023)\citenamefont {Lechner}, \citenamefont {Pennetta}, \citenamefont {Blaha}, \citenamefont {Schneeweiss}, \citenamefont {Rauschenbeutel},\ and\ \citenamefont {Volz}}]{Arno-delay2023}%
  \BibitemOpen
  \bibfield  {author} {\bibinfo {author} {\bibfnamefont {D.}~\bibnamefont {Lechner}}, \bibinfo {author} {\bibfnamefont {R.}~\bibnamefont {Pennetta}}, \bibinfo {author} {\bibfnamefont {M.}~\bibnamefont {Blaha}}, \bibinfo {author} {\bibfnamefont {P.}~\bibnamefont {Schneeweiss}}, \bibinfo {author} {\bibfnamefont {A.}~\bibnamefont {Rauschenbeutel}},\ and\ \bibinfo {author} {\bibfnamefont {J.}~\bibnamefont {Volz}},\ }\bibfield  {title} {\emph {\bibinfo {title} {Light-matter interaction at the transition between cavity and waveguide qed}},\ }\href {https://doi.org/10.1103/PhysRevLett.131.103603} {\bibfield  {journal} {\bibinfo  {journal} {Phys. Rev. Lett.}\ }\textbf {\bibinfo {volume} {131}},\ \bibinfo {pages} {103603} (\bibinfo {year} {2023})}\BibitemShut {NoStop}%
\bibitem [{\citenamefont {Zheng}\ and\ \citenamefont {Baranger}(2013)}]{ZhengPRL13}%
  \BibitemOpen
  \bibfield  {author} {\bibinfo {author} {\bibfnamefont {H.}~\bibnamefont {Zheng}}\ and\ \bibinfo {author} {\bibfnamefont {H.~U.}\ \bibnamefont {Baranger}},\ }\bibfield  {title} {\emph {\bibinfo {title} {{Persistent quantum beats and long-distance entanglement from waveguide-mediated interactions}}},\ }\href {https://doi.org/10.1103/PhysRevLett.110.113601} {\bibfield  {journal} {\bibinfo  {journal} {Physical Review Letters}\ }\textbf {\bibinfo {volume} {110}},\ \bibinfo {pages} {113601} (\bibinfo {year} {2013})}\BibitemShut {NoStop}%
\bibitem [{\citenamefont {Carmele}\ \emph {et~al.}(2013)\citenamefont {Carmele}, \citenamefont {Kabuss}, \citenamefont {Schulze}, \citenamefont {Reitzenstein},\ and\ \citenamefont {Knorr}}]{CarmelePRL13}%
  \BibitemOpen
  \bibfield  {author} {\bibinfo {author} {\bibfnamefont {A.}~\bibnamefont {Carmele}}, \bibinfo {author} {\bibfnamefont {J.}~\bibnamefont {Kabuss}}, \bibinfo {author} {\bibfnamefont {F.}~\bibnamefont {Schulze}}, \bibinfo {author} {\bibfnamefont {S.}~\bibnamefont {Reitzenstein}},\ and\ \bibinfo {author} {\bibfnamefont {A.}~\bibnamefont {Knorr}},\ }\bibfield  {title} {\emph {\bibinfo {title} {{Single photon delayed feedback: A way to stabilize intrinsic quantum cavity electrodynamics}}},\ }\href {https://doi.org/10.1103/PhysRevLett.110.013601} {\bibfield  {journal} {\bibinfo  {journal} {Physical Review Letters}\ }\textbf {\bibinfo {volume} {110}},\ \bibinfo {pages} {13601} (\bibinfo {year} {2013})}\BibitemShut {NoStop}%
\bibitem [{\citenamefont {Kabuss}\ \emph {et~al.}(2015)\citenamefont {Kabuss}, \citenamefont {Krimer}, \citenamefont {Rotter}, \citenamefont {Stannigel}, \citenamefont {Knorr},\ and\ \citenamefont {Carmele}}]{CarmelePRA13}%
  \BibitemOpen
  \bibfield  {author} {\bibinfo {author} {\bibfnamefont {J.}~\bibnamefont {Kabuss}}, \bibinfo {author} {\bibfnamefont {D.~O.}\ \bibnamefont {Krimer}}, \bibinfo {author} {\bibfnamefont {S.}~\bibnamefont {Rotter}}, \bibinfo {author} {\bibfnamefont {K.}~\bibnamefont {Stannigel}}, \bibinfo {author} {\bibfnamefont {A.}~\bibnamefont {Knorr}},\ and\ \bibinfo {author} {\bibfnamefont {A.}~\bibnamefont {Carmele}},\ }\bibfield  {title} {\emph {\bibinfo {title} {Analytical study of quantum-feedback-enhanced rabi oscillations}},\ }\href {https://doi.org/10.1103/PhysRevA.92.053801} {\bibfield  {journal} {\bibinfo  {journal} {Phys. Rev. A}\ }\textbf {\bibinfo {volume} {92}},\ \bibinfo {pages} {053801} (\bibinfo {year} {2015})}\BibitemShut {NoStop}%
\bibitem [{\citenamefont {Grimsmo}(2015)}]{GrimsmoPRL15}%
  \BibitemOpen
  \bibfield  {author} {\bibinfo {author} {\bibfnamefont {A.~L.}\ \bibnamefont {Grimsmo}},\ }\bibfield  {title} {\emph {\bibinfo {title} {{Time-Delayed Quantum Feedback Control}}},\ }\href {https://doi.org/10.1103/PhysRevLett.115.060402} {\bibfield  {journal} {\bibinfo  {journal} {Physical Review Letters}\ }\textbf {\bibinfo {volume} {115}},\ \bibinfo {pages} {60402} (\bibinfo {year} {2015})}\BibitemShut {NoStop}%
\bibitem [{\citenamefont {Laakso}\ and\ \citenamefont {Pletyukhov}(2014)}]{LaaksoPRL14}%
  \BibitemOpen
  \bibfield  {author} {\bibinfo {author} {\bibfnamefont {M.}~\bibnamefont {Laakso}}\ and\ \bibinfo {author} {\bibfnamefont {M.}~\bibnamefont {Pletyukhov}},\ }\bibfield  {title} {\emph {\bibinfo {title} {{Scattering of two photons from two distant qubits: Exact solution}}},\ }\href {https://doi.org/10.1103/PhysRevLett.113.183601} {\bibfield  {journal} {\bibinfo  {journal} {Physical Review Letters}\ }\textbf {\bibinfo {volume} {113}},\ \bibinfo {pages} {183601} (\bibinfo {year} {2014})}\BibitemShut {NoStop}%
\bibitem [{\citenamefont {Fang}\ and\ \citenamefont {Baranger}(2015)}]{FangPRA15}%
  \BibitemOpen
  \bibfield  {author} {\bibinfo {author} {\bibfnamefont {Y.~L.~L.}\ \bibnamefont {Fang}}\ and\ \bibinfo {author} {\bibfnamefont {H.~U.}\ \bibnamefont {Baranger}},\ }\bibfield  {title} {\emph {\bibinfo {title} {{Waveguide QED: Power spectra and correlations of two photons scattered off multiple distant qubits and a mirror}}},\ }\href {https://doi.org/10.1103/PhysRevA.91.053845} {\bibfield  {journal} {\bibinfo  {journal} {Physical Review A}\ }\textbf {\bibinfo {volume} {91}},\ \bibinfo {pages} {53845} (\bibinfo {year} {2015})},\ \Eprint {https://arxiv.org/abs/1502.03803} {arXiv:1502.03803} \BibitemShut {NoStop}%
\bibitem [{\citenamefont {Pichler}\ \emph {et~al.}(2017)\citenamefont {Pichler}, \citenamefont {Choi}, \citenamefont {Zoller},\ and\ \citenamefont {Lukin}}]{PichlerPNAS17}%
  \BibitemOpen
  \bibfield  {author} {\bibinfo {author} {\bibfnamefont {H.}~\bibnamefont {Pichler}}, \bibinfo {author} {\bibfnamefont {S.}~\bibnamefont {Choi}}, \bibinfo {author} {\bibfnamefont {P.}~\bibnamefont {Zoller}},\ and\ \bibinfo {author} {\bibfnamefont {M.~D.}\ \bibnamefont {Lukin}},\ }\bibfield  {title} {\emph {\bibinfo {title} {{Universal photonic quantum computation via time-delayed feedback}}},\ }\href {https://doi.org/10.1073/pnas.1711003114} {\bibfield  {journal} {\bibinfo  {journal} {Proceedings of the National Academy of Sciences}\ }\textbf {\bibinfo {volume} {114}},\ \bibinfo {pages} {11362--11367} (\bibinfo {year} {2017})}\BibitemShut {NoStop}%
\bibitem [{\citenamefont {Calaj{\'{o}}}\ \emph {et~al.}(2019)\citenamefont {Calaj{\'{o}}}, \citenamefont {Fang}, \citenamefont {Baranger},\ and\ \citenamefont {Ciccarello}}]{Calajo2019}%
  \BibitemOpen
  \bibfield  {author} {\bibinfo {author} {\bibfnamefont {G.}~\bibnamefont {Calaj{\'{o}}}}, \bibinfo {author} {\bibfnamefont {Y.-L.~L.}\ \bibnamefont {Fang}}, \bibinfo {author} {\bibfnamefont {H.~U.}\ \bibnamefont {Baranger}},\ and\ \bibinfo {author} {\bibfnamefont {F.}~\bibnamefont {Ciccarello}},\ }\bibfield  {title} {\emph {\bibinfo {title} {{Exciting a Bound State in the Continuum through Multiphoton Scattering Plus Delayed Quantum Feedback}}},\ }\href {https://doi.org/10.1103/PhysRevLett.122.073601} {\bibfield  {journal} {\bibinfo  {journal} {Physical Review Letters}\ }\textbf {\bibinfo {volume} {122}},\ \bibinfo {pages} {073601} (\bibinfo {year} {2019})}\BibitemShut {NoStop}%
\bibitem [{\citenamefont {Trivedi}\ \emph {et~al.}(2021{\natexlab{a}})\citenamefont {Trivedi}, \citenamefont {Malz}, \citenamefont {Sun}, \citenamefont {Fan},\ and\ \citenamefont {Vu\ifmmode \check{c}\else \v{c}\fi{}kovi\ifmmode~\acute{c}\else \'{c}\fi{}}}]{TrivediPRA21}%
  \BibitemOpen
  \bibfield  {author} {\bibinfo {author} {\bibfnamefont {R.}~\bibnamefont {Trivedi}}, \bibinfo {author} {\bibfnamefont {D.}~\bibnamefont {Malz}}, \bibinfo {author} {\bibfnamefont {S.}~\bibnamefont {Sun}}, \bibinfo {author} {\bibfnamefont {S.}~\bibnamefont {Fan}},\ and\ \bibinfo {author} {\bibfnamefont {J.}~\bibnamefont {Vu\ifmmode \check{c}\else \v{c}\fi{}kovi\ifmmode~\acute{c}\else \'{c}\fi{}}},\ }\bibfield  {title} {\emph {\bibinfo {title} {Optimal two-photon excitation of bound states in non-markovian waveguide qed}},\ }\href {https://doi.org/10.1103/PhysRevA.104.013705} {\bibfield  {journal} {\bibinfo  {journal} {Phys. Rev. A}\ }\textbf {\bibinfo {volume} {104}},\ \bibinfo {pages} {013705} (\bibinfo {year} {2021}{\natexlab{a}})}\BibitemShut {NoStop}%
\bibitem [{\citenamefont {Barkemeyer}\ \emph {et~al.}(2021{\natexlab{a}})\citenamefont {Barkemeyer}, \citenamefont {Knorr},\ and\ \citenamefont {Carmele}}]{CarmeleBIC21}%
  \BibitemOpen
  \bibfield  {author} {\bibinfo {author} {\bibfnamefont {K.}~\bibnamefont {Barkemeyer}}, \bibinfo {author} {\bibfnamefont {A.}~\bibnamefont {Knorr}},\ and\ \bibinfo {author} {\bibfnamefont {A.}~\bibnamefont {Carmele}},\ }\bibfield  {title} {\emph {\bibinfo {title} {Strongly entangled system-reservoir dynamics with multiphoton pulses beyond the two-excitation limit: Exciting the atom-photon bound state}},\ }\href {https://doi.org/10.1103/PhysRevA.103.033704} {\bibfield  {journal} {\bibinfo  {journal} {Phys. Rev. A}\ }\textbf {\bibinfo {volume} {103}},\ \bibinfo {pages} {033704} (\bibinfo {year} {2021}{\natexlab{a}})}\BibitemShut {NoStop}%
\bibitem [{\citenamefont {Sinha}\ \emph {et~al.}(2020{\natexlab{a}})\citenamefont {Sinha}, \citenamefont {Meystre}, \citenamefont {Goldschmidt}, \citenamefont {Fatemi}, \citenamefont {Rolston},\ and\ \citenamefont {Solano}}]{SinhaPRL20}%
  \BibitemOpen
  \bibfield  {author} {\bibinfo {author} {\bibfnamefont {K.}~\bibnamefont {Sinha}}, \bibinfo {author} {\bibfnamefont {P.}~\bibnamefont {Meystre}}, \bibinfo {author} {\bibfnamefont {E.~A.}\ \bibnamefont {Goldschmidt}}, \bibinfo {author} {\bibfnamefont {F.~K.}\ \bibnamefont {Fatemi}}, \bibinfo {author} {\bibfnamefont {S.~L.}\ \bibnamefont {Rolston}},\ and\ \bibinfo {author} {\bibfnamefont {P.}~\bibnamefont {Solano}},\ }\bibfield  {title} {\emph {\bibinfo {title} {Non-markovian collective emission from macroscopically separated emitters}},\ }\href {https://doi.org/10.1103/PhysRevLett.124.043603} {\bibfield  {journal} {\bibinfo  {journal} {Phys. Rev. Lett.}\ }\textbf {\bibinfo {volume} {124}},\ \bibinfo {pages} {043603} (\bibinfo {year} {2020}{\natexlab{a}})}\BibitemShut {NoStop}%
\bibitem [{\citenamefont {Dinc}\ and\ \citenamefont {Bra\ifmmode~\acute{n}\else \'{n}\fi{}czyk}(2019)}]{DincPRR19}%
  \BibitemOpen
  \bibfield  {author} {\bibinfo {author} {\bibfnamefont {F.}~\bibnamefont {Dinc}}\ and\ \bibinfo {author} {\bibfnamefont {A.~M.}\ \bibnamefont {Bra\ifmmode~\acute{n}\else \'{n}\fi{}czyk}},\ }\bibfield  {title} {\emph {\bibinfo {title} {Non-markovian super-superradiance in a linear chain of up to 100 qubits}},\ }\href {https://doi.org/10.1103/PhysRevResearch.1.032042} {\bibfield  {journal} {\bibinfo  {journal} {Phys. Rev. Res.}\ }\textbf {\bibinfo {volume} {1}},\ \bibinfo {pages} {032042} (\bibinfo {year} {2019})}\BibitemShut {NoStop}%
\bibitem [{\citenamefont {Sinha}\ \emph {et~al.}(2020{\natexlab{b}})\citenamefont {Sinha}, \citenamefont {Gonz\'alez-Tudela}, \citenamefont {Lu},\ and\ \citenamefont {Solano}}]{SinhaPRA20}%
  \BibitemOpen
  \bibfield  {author} {\bibinfo {author} {\bibfnamefont {K.}~\bibnamefont {Sinha}}, \bibinfo {author} {\bibfnamefont {A.}~\bibnamefont {Gonz\'alez-Tudela}}, \bibinfo {author} {\bibfnamefont {Y.}~\bibnamefont {Lu}},\ and\ \bibinfo {author} {\bibfnamefont {P.}~\bibnamefont {Solano}},\ }\bibfield  {title} {\emph {\bibinfo {title} {Collective radiation from distant emitters}},\ }\href {https://doi.org/10.1103/PhysRevA.102.043718} {\bibfield  {journal} {\bibinfo  {journal} {Phys. Rev. A}\ }\textbf {\bibinfo {volume} {102}},\ \bibinfo {pages} {043718} (\bibinfo {year} {2020}{\natexlab{b}})}\BibitemShut {NoStop}%
\bibitem [{\citenamefont {Gonz\'alez-Guti\'errez}\ \emph {et~al.}(2021)\citenamefont {Gonz\'alez-Guti\'errez}, \citenamefont {Rom\'an-Roche},\ and\ \citenamefont {Zueco}}]{ZuecoPRA21}%
  \BibitemOpen
  \bibfield  {author} {\bibinfo {author} {\bibfnamefont {C.~A.}\ \bibnamefont {Gonz\'alez-Guti\'errez}}, \bibinfo {author} {\bibfnamefont {J.}~\bibnamefont {Rom\'an-Roche}},\ and\ \bibinfo {author} {\bibfnamefont {D.}~\bibnamefont {Zueco}},\ }\bibfield  {title} {\emph {\bibinfo {title} {Distant emitters in ultrastrong waveguide qed: Ground-state properties and non-markovian dynamics}},\ }\href {https://doi.org/10.1103/PhysRevA.104.053701} {\bibfield  {journal} {\bibinfo  {journal} {Phys. Rev. A}\ }\textbf {\bibinfo {volume} {104}},\ \bibinfo {pages} {053701} (\bibinfo {year} {2021})}\BibitemShut {NoStop}%
\bibitem [{\citenamefont {Carmele}\ \emph {et~al.}(2020)\citenamefont {Carmele}, \citenamefont {Nemet}, \citenamefont {Canela},\ and\ \citenamefont {Parkins}}]{CarmelePRR20}%
  \BibitemOpen
  \bibfield  {author} {\bibinfo {author} {\bibfnamefont {A.}~\bibnamefont {Carmele}}, \bibinfo {author} {\bibfnamefont {N.}~\bibnamefont {Nemet}}, \bibinfo {author} {\bibfnamefont {V.}~\bibnamefont {Canela}},\ and\ \bibinfo {author} {\bibfnamefont {S.}~\bibnamefont {Parkins}},\ }\bibfield  {title} {\emph {\bibinfo {title} {Pronounced non-markovian features in multiply excited, multiple emitter waveguide qed: Retardation induced anomalous population trapping}},\ }\href {https://doi.org/10.1103/PhysRevResearch.2.013238} {\bibfield  {journal} {\bibinfo  {journal} {Phys. Rev. Res.}\ }\textbf {\bibinfo {volume} {2}},\ \bibinfo {pages} {013238} (\bibinfo {year} {2020})}\BibitemShut {NoStop}%
\bibitem [{\citenamefont {Guo}\ \emph {et~al.}(2020)\citenamefont {Guo}, \citenamefont {Kockum}, \citenamefont {Marquardt},\ and\ \citenamefont {Johansson}}]{KockumPRR20}%
  \BibitemOpen
  \bibfield  {author} {\bibinfo {author} {\bibfnamefont {L.}~\bibnamefont {Guo}}, \bibinfo {author} {\bibfnamefont {A.~F.}\ \bibnamefont {Kockum}}, \bibinfo {author} {\bibfnamefont {F.}~\bibnamefont {Marquardt}},\ and\ \bibinfo {author} {\bibfnamefont {G.}~\bibnamefont {Johansson}},\ }\bibfield  {title} {\emph {\bibinfo {title} {Oscillating bound states for a giant atom}},\ }\href {https://doi.org/10.1103/PhysRevResearch.2.043014} {\bibfield  {journal} {\bibinfo  {journal} {Phys. Rev. Research}\ }\textbf {\bibinfo {volume} {2}},\ \bibinfo {pages} {043014} (\bibinfo {year} {2020})}\BibitemShut {NoStop}%
\bibitem [{\citenamefont {Barkemeyer}\ \emph {et~al.}(2021{\natexlab{b}})\citenamefont {Barkemeyer}, \citenamefont {Hohn}, \citenamefont {Reitzenstein},\ and\ \citenamefont {Carmele}}]{CarmeleET21}%
  \BibitemOpen
  \bibfield  {author} {\bibinfo {author} {\bibfnamefont {K.}~\bibnamefont {Barkemeyer}}, \bibinfo {author} {\bibfnamefont {M.}~\bibnamefont {Hohn}}, \bibinfo {author} {\bibfnamefont {S.}~\bibnamefont {Reitzenstein}},\ and\ \bibinfo {author} {\bibfnamefont {A.}~\bibnamefont {Carmele}},\ }\bibfield  {title} {\emph {\bibinfo {title} {Boosting energy-time entanglement using coherent time-delayed feedback}},\ }\href {https://doi.org/10.1103/PhysRevA.103.062423} {\bibfield  {journal} {\bibinfo  {journal} {Phys. Rev. A}\ }\textbf {\bibinfo {volume} {103}},\ \bibinfo {pages} {062423} (\bibinfo {year} {2021}{\natexlab{b}})}\BibitemShut {NoStop}%
\bibitem [{\citenamefont {Crowder}\ \emph {et~al.}(2023)\citenamefont {Crowder}, \citenamefont {Ramunno},\ and\ \citenamefont {Hughes}}]{crowder2023improving}%
  \BibitemOpen
  \bibfield  {author} {\bibinfo {author} {\bibfnamefont {G.}~\bibnamefont {Crowder}}, \bibinfo {author} {\bibfnamefont {L.}~\bibnamefont {Ramunno}},\ and\ \bibinfo {author} {\bibfnamefont {S.}~\bibnamefont {Hughes}},\ }\href@noop {} {\bibinfo {title} {Improving on-demand single photon source coherence and indistinguishability through a time-delayed coherent feedback}} (\bibinfo {year} {2023}),\ \Eprint {https://arxiv.org/abs/2302.08093} {arXiv:2302.08093 [quant-ph]} \BibitemShut {NoStop}%
\bibitem [{\citenamefont {Ask}\ and\ \citenamefont {Johansson}(2022)}]{AskPRL22}%
  \BibitemOpen
  \bibfield  {author} {\bibinfo {author} {\bibfnamefont {A.}~\bibnamefont {Ask}}\ and\ \bibinfo {author} {\bibfnamefont {G.}~\bibnamefont {Johansson}},\ }\bibfield  {title} {\emph {\bibinfo {title} {Non-markovian steady states of a driven two-level system}},\ }\href {https://doi.org/10.1103/PhysRevLett.128.083603} {\bibfield  {journal} {\bibinfo  {journal} {Phys. Rev. Lett.}\ }\textbf {\bibinfo {volume} {128}},\ \bibinfo {pages} {083603} (\bibinfo {year} {2022})}\BibitemShut {NoStop}%
\bibitem [{\citenamefont {Pichler}\ and\ \citenamefont {Zoller}(2016)}]{pichlerPhotonic2016}%
  \BibitemOpen
  \bibfield  {author} {\bibinfo {author} {\bibfnamefont {H.}~\bibnamefont {Pichler}}\ and\ \bibinfo {author} {\bibfnamefont {P.}~\bibnamefont {Zoller}},\ }\bibfield  {title} {\emph {\bibinfo {title} {{Photonic Circuits with Time Delays and Quantum Feedback}}},\ }\href {https://doi.org/10.1103/PhysRevLett.116.093601} {\bibfield  {journal} {\bibinfo  {journal} {Physical Review Letters}\ }\textbf {\bibinfo {volume} {116}},\ \bibinfo {pages} {93601} (\bibinfo {year} {2016})}\BibitemShut {NoStop}%
\bibitem [{\citenamefont {Ramos}\ \emph {et~al.}(2016)\citenamefont {Ramos}, \citenamefont {Vermersch}, \citenamefont {Hauke}, \citenamefont {Pichler},\ and\ \citenamefont {Zoller}}]{RamosPRA16}%
  \BibitemOpen
  \bibfield  {author} {\bibinfo {author} {\bibfnamefont {T.}~\bibnamefont {Ramos}}, \bibinfo {author} {\bibfnamefont {B.}~\bibnamefont {Vermersch}}, \bibinfo {author} {\bibfnamefont {P.}~\bibnamefont {Hauke}}, \bibinfo {author} {\bibfnamefont {H.}~\bibnamefont {Pichler}},\ and\ \bibinfo {author} {\bibfnamefont {P.}~\bibnamefont {Zoller}},\ }\bibfield  {title} {\emph {\bibinfo {title} {Non-markovian dynamics in chiral quantum networks with spins and photons}},\ }\href {https://doi.org/10.1103/PhysRevA.93.062104} {\bibfield  {journal} {\bibinfo  {journal} {Phys. Rev. A}\ }\textbf {\bibinfo {volume} {93}},\ \bibinfo {pages} {062104} (\bibinfo {year} {2016})}\BibitemShut {NoStop}%
\bibitem [{\citenamefont {Trivedi}\ \emph {et~al.}(2019)\citenamefont {Trivedi}, \citenamefont {Fischer}, \citenamefont {Mishra},\ and\ \citenamefont {Vu\ifmmode \check{c}\else \v{c}\fi{}kovi\ifmmode~\acute{c}\else \'{c}\fi{}}}]{trivediPRA19}%
  \BibitemOpen
  \bibfield  {author} {\bibinfo {author} {\bibfnamefont {R.}~\bibnamefont {Trivedi}}, \bibinfo {author} {\bibfnamefont {K.}~\bibnamefont {Fischer}}, \bibinfo {author} {\bibfnamefont {S.~D.}\ \bibnamefont {Mishra}},\ and\ \bibinfo {author} {\bibfnamefont {J.}~\bibnamefont {Vu\ifmmode \check{c}\else \v{c}\fi{}kovi\ifmmode~\acute{c}\else \'{c}\fi{}}},\ }\bibfield  {title} {\emph {\bibinfo {title} {Point-coupling hamiltonian for frequency-independent linear optical devices}},\ }\href {https://doi.org/10.1103/PhysRevA.100.043827} {\bibfield  {journal} {\bibinfo  {journal} {Phys. Rev. A}\ }\textbf {\bibinfo {volume} {100}},\ \bibinfo {pages} {043827} (\bibinfo {year} {2019})}\BibitemShut {NoStop}%
\bibitem [{\citenamefont {Whalen}(2019)}]{WhalenPRA19}%
  \BibitemOpen
  \bibfield  {author} {\bibinfo {author} {\bibfnamefont {S.~J.}\ \bibnamefont {Whalen}},\ }\bibfield  {title} {\emph {\bibinfo {title} {Collision model for non-markovian quantum trajectories}},\ }\href {https://doi.org/10.1103/PhysRevA.100.052113} {\bibfield  {journal} {\bibinfo  {journal} {Phys. Rev. A}\ }\textbf {\bibinfo {volume} {100}},\ \bibinfo {pages} {052113} (\bibinfo {year} {2019})}\BibitemShut {NoStop}%
\bibitem [{\citenamefont {Fang}(2019)}]{LeoCP19}%
  \BibitemOpen
  \bibfield  {author} {\bibinfo {author} {\bibfnamefont {Y.-L.~L.}\ \bibnamefont {Fang}},\ }\bibfield  {title} {\emph {\bibinfo {title} {Fdtd: Solving 1+1d delay pde in parallel}},\ }\href {https://doi.org/https://doi.org/10.1016/j.cpc.2018.08.018} {\bibfield  {journal} {\bibinfo  {journal} {Computer Physics Communications}\ }\textbf {\bibinfo {volume} {235}},\ \bibinfo {pages} {422--432} (\bibinfo {year} {2019})}\BibitemShut {NoStop}%
\bibitem [{\citenamefont {Dinc}\ \emph {et~al.}(2019)\citenamefont {Dinc}, \citenamefont {Ercan},\ and\ \citenamefont {Bra{\'{n}}czyk}}]{dinc2019exact}%
  \BibitemOpen
  \bibfield  {author} {\bibinfo {author} {\bibfnamefont {F.}~\bibnamefont {Dinc}}, \bibinfo {author} {\bibfnamefont {{\.{I}}.}~\bibnamefont {Ercan}},\ and\ \bibinfo {author} {\bibfnamefont {A.~M.}\ \bibnamefont {Bra{\'{n}}czyk}},\ }\bibfield  {title} {\emph {\bibinfo {title} {Exact {M}arkovian and non-{M}arkovian time dynamics in waveguide {QED}: collective interactions, bound states in continuum, superradiance and subradiance}},\ }\href {https://doi.org/10.22331/q-2019-12-09-213} {\bibfield  {journal} {\bibinfo  {journal} {{Quantum}}\ }\textbf {\bibinfo {volume} {3}},\ \bibinfo {pages} {213} (\bibinfo {year} {2019})}\BibitemShut {NoStop}%
\bibitem [{\citenamefont {Crowder}\ \emph {et~al.}(2020)\citenamefont {Crowder}, \citenamefont {Carmichael},\ and\ \citenamefont {Hughes}}]{CrowderPRA20}%
  \BibitemOpen
  \bibfield  {author} {\bibinfo {author} {\bibfnamefont {G.}~\bibnamefont {Crowder}}, \bibinfo {author} {\bibfnamefont {H.}~\bibnamefont {Carmichael}},\ and\ \bibinfo {author} {\bibfnamefont {S.}~\bibnamefont {Hughes}},\ }\bibfield  {title} {\emph {\bibinfo {title} {Quantum trajectory theory of few-photon cavity-qed systems with a time-delayed coherent feedback}},\ }\href {https://doi.org/10.1103/PhysRevA.101.023807} {\bibfield  {journal} {\bibinfo  {journal} {Phys. Rev. A}\ }\textbf {\bibinfo {volume} {101}},\ \bibinfo {pages} {023807} (\bibinfo {year} {2020})}\BibitemShut {NoStop}%
\bibitem [{\citenamefont {Arranz~Regidor}\ \emph {et~al.}(2021)\citenamefont {Arranz~Regidor}, \citenamefont {Crowder}, \citenamefont {Carmichael},\ and\ \citenamefont {Hughes}}]{HughesPRR21}%
  \BibitemOpen
  \bibfield  {author} {\bibinfo {author} {\bibfnamefont {S.}~\bibnamefont {Arranz~Regidor}}, \bibinfo {author} {\bibfnamefont {G.}~\bibnamefont {Crowder}}, \bibinfo {author} {\bibfnamefont {H.}~\bibnamefont {Carmichael}},\ and\ \bibinfo {author} {\bibfnamefont {S.}~\bibnamefont {Hughes}},\ }\bibfield  {title} {\emph {\bibinfo {title} {Modeling quantum light-matter interactions in waveguide qed with retardation, nonlinear interactions, and a time-delayed feedback: Matrix product states versus a space-discretized waveguide model}},\ }\href {https://doi.org/10.1103/PhysRevResearch.3.023030} {\bibfield  {journal} {\bibinfo  {journal} {Phys. Rev. Res.}\ }\textbf {\bibinfo {volume} {3}},\ \bibinfo {pages} {023030} (\bibinfo {year} {2021})}\BibitemShut {NoStop}%
\bibitem [{\citenamefont {Crowder}\ \emph {et~al.}(2022)\citenamefont {Crowder}, \citenamefont {Ramunno},\ and\ \citenamefont {Hughes}}]{CrowderPRA22}%
  \BibitemOpen
  \bibfield  {author} {\bibinfo {author} {\bibfnamefont {G.}~\bibnamefont {Crowder}}, \bibinfo {author} {\bibfnamefont {L.}~\bibnamefont {Ramunno}},\ and\ \bibinfo {author} {\bibfnamefont {S.}~\bibnamefont {Hughes}},\ }\bibfield  {title} {\emph {\bibinfo {title} {Quantum trajectory theory and simulations of nonlinear spectra and multiphoton effects in waveguide-qed systems with a time-delayed coherent feedback}},\ }\href {https://doi.org/10.1103/PhysRevA.106.013714} {\bibfield  {journal} {\bibinfo  {journal} {Phys. Rev. A}\ }\textbf {\bibinfo {volume} {106}},\ \bibinfo {pages} {013714} (\bibinfo {year} {2022})}\BibitemShut {NoStop}%
\bibitem [{\citenamefont {Zhang}\ \emph {et~al.}(2022)\citenamefont {Zhang}, \citenamefont {Klapp},\ and\ \citenamefont {Metelmann}}]{zhang2022embedding}%
  \BibitemOpen
  \bibfield  {author} {\bibinfo {author} {\bibfnamefont {X.~H.}\ \bibnamefont {Zhang}}, \bibinfo {author} {\bibfnamefont {S.}~\bibnamefont {Klapp}},\ and\ \bibinfo {author} {\bibfnamefont {A.}~\bibnamefont {Metelmann}},\ }\bibfield  {title} {\emph {\bibinfo {title} {Embedding of time-delayed quantum feedback in a nonreciprocal array}},\ }\href@noop {} {\bibfield  {journal} {\bibinfo  {journal} {arXiv preprint arXiv:2204.02367}\ } (\bibinfo {year} {2022})}\BibitemShut {NoStop}%
\bibitem [{\citenamefont {Barkemeyer}\ \emph {et~al.}(2022)\citenamefont {Barkemeyer}, \citenamefont {Knorr},\ and\ \citenamefont {Carmele}}]{CarmelePRA22}%
  \BibitemOpen
  \bibfield  {author} {\bibinfo {author} {\bibfnamefont {K.}~\bibnamefont {Barkemeyer}}, \bibinfo {author} {\bibfnamefont {A.}~\bibnamefont {Knorr}},\ and\ \bibinfo {author} {\bibfnamefont {A.}~\bibnamefont {Carmele}},\ }\bibfield  {title} {\emph {\bibinfo {title} {Heisenberg treatment of multiphoton pulses in waveguide qed with time-delayed feedback}},\ }\href {https://doi.org/10.1103/PhysRevA.106.023708} {\bibfield  {journal} {\bibinfo  {journal} {Phys. Rev. A}\ }\textbf {\bibinfo {volume} {106}},\ \bibinfo {pages} {023708} (\bibinfo {year} {2022})}\BibitemShut {NoStop}%
\bibitem [{\citenamefont {Vodenkova}\ and\ \citenamefont {Pichler}()}]{vodenkova2023continuous}%
  \BibitemOpen
  \bibfield  {author} {\bibinfo {author} {\bibfnamefont {K.}~\bibnamefont {Vodenkova}}\ and\ \bibinfo {author} {\bibfnamefont {H.}~\bibnamefont {Pichler}},\ }\bibfield  {title} {\emph {\bibinfo {title} {Continuous coherent quantum feedback with time delays: Tensor network solution}},\ }\href@noop {} {\bibinfo  {journal} {arXiv:2311.07302}\ }\BibitemShut {NoStop}%
\bibitem [{\citenamefont {Dinc}(2020)}]{DincPRA19}%
  \BibitemOpen
\bibfield  {journal} {  }\bibfield  {author} {\bibinfo {author} {\bibfnamefont {F.}~\bibnamefont {Dinc}},\ }\bibfield  {title} {\emph {\bibinfo {title} {Diagrammatic approach for analytical non-markovian time evolution: Fermi's two-atom problem and causality in waveguide quantum electrodynamics}},\ }\href {https://doi.org/10.1103/PhysRevA.102.013727} {\bibfield  {journal} {\bibinfo  {journal} {Phys. Rev. A}\ }\textbf {\bibinfo {volume} {102}},\ \bibinfo {pages} {013727} (\bibinfo {year} {2020})}\BibitemShut {NoStop}%
\bibitem [{\citenamefont {Piasotski}\ and\ \citenamefont {Pletyukhov}(2021)}]{PletyuPRA21}%
  \BibitemOpen
  \bibfield  {author} {\bibinfo {author} {\bibfnamefont {K.}~\bibnamefont {Piasotski}}\ and\ \bibinfo {author} {\bibfnamefont {M.}~\bibnamefont {Pletyukhov}},\ }\bibfield  {title} {\emph {\bibinfo {title} {Diagrammatic approach to scattering of multiphoton states in waveguide qed}},\ }\href {https://doi.org/10.1103/PhysRevA.104.023709} {\bibfield  {journal} {\bibinfo  {journal} {Phys. Rev. A}\ }\textbf {\bibinfo {volume} {104}},\ \bibinfo {pages} {023709} (\bibinfo {year} {2021})}\BibitemShut {NoStop}%
\bibitem [{\citenamefont {Milonni}\ \emph {et~al.}(1983)\citenamefont {Milonni}, \citenamefont {Ackerhalt}, \citenamefont {Galbraith},\ and\ \citenamefont {Shih}}]{MilonniPRA83}%
  \BibitemOpen
  \bibfield  {author} {\bibinfo {author} {\bibfnamefont {P.~W.}\ \bibnamefont {Milonni}}, \bibinfo {author} {\bibfnamefont {J.~R.}\ \bibnamefont {Ackerhalt}}, \bibinfo {author} {\bibfnamefont {H.~W.}\ \bibnamefont {Galbraith}},\ and\ \bibinfo {author} {\bibfnamefont {M.-L.}\ \bibnamefont {Shih}},\ }\bibfield  {title} {\emph {\bibinfo {title} {Exponential decay, recurrences, and quantum-mechanical spreading in a quasicontinuum model}},\ }\href {https://doi.org/10.1103/PhysRevA.28.32} {\bibfield  {journal} {\bibinfo  {journal} {Phys. Rev. A}\ }\textbf {\bibinfo {volume} {28}},\ \bibinfo {pages} {32--39} (\bibinfo {year} {1983})}\BibitemShut {NoStop}%
\bibitem [{\citenamefont {Cook}\ and\ \citenamefont {Milonni}(1987)}]{CookPRA87}%
  \BibitemOpen
  \bibfield  {author} {\bibinfo {author} {\bibfnamefont {R.~J.}\ \bibnamefont {Cook}}\ and\ \bibinfo {author} {\bibfnamefont {P.~W.}\ \bibnamefont {Milonni}},\ }\bibfield  {title} {\emph {\bibinfo {title} {{Quantum theory of an atom near partially reflecting walls}}},\ }\href {https://doi.org/10.1103/PhysRevA.35.5081} {\bibfield  {journal} {\bibinfo  {journal} {Physical Review A}\ }\textbf {\bibinfo {volume} {35}},\ \bibinfo {pages} {5081--5087} (\bibinfo {year} {1987})}\BibitemShut {NoStop}%
\bibitem [{\citenamefont {Gießen}\ \emph {et~al.}(1996)\citenamefont {Gießen}, \citenamefont {Berger}, \citenamefont {Mohs}, \citenamefont {Meystre},\ and\ \citenamefont {Yelin}}]{gie1996cavity}%
  \BibitemOpen
  \bibfield  {author} {\bibinfo {author} {\bibfnamefont {H.}~\bibnamefont {Gießen}}, \bibinfo {author} {\bibfnamefont {J.~D.}\ \bibnamefont {Berger}}, \bibinfo {author} {\bibfnamefont {G.}~\bibnamefont {Mohs}}, \bibinfo {author} {\bibfnamefont {P.}~\bibnamefont {Meystre}},\ and\ \bibinfo {author} {\bibfnamefont {S.~F.}\ \bibnamefont {Yelin}},\ }\bibfield  {title} {\emph {\bibinfo {title} {Cavity-modified spontaneous emission: From rabi oscillations to exponential decay}},\ }\href {https://doi.org/10.1103/PhysRevA.53.2816} {\bibfield  {journal} {\bibinfo  {journal} {Phys. Rev. A}\ }\textbf {\bibinfo {volume} {53}},\ \bibinfo {pages} {2816--2821} (\bibinfo {year} {1996})}\BibitemShut {NoStop}%
\bibitem [{\citenamefont {Krimer}\ \emph {et~al.}(2014)\citenamefont {Krimer}, \citenamefont {Liertzer}, \citenamefont {Rotter},\ and\ \citenamefont {T\"ureci}}]{RotterVolterra14}%
  \BibitemOpen
  \bibfield  {author} {\bibinfo {author} {\bibfnamefont {D.~O.}\ \bibnamefont {Krimer}}, \bibinfo {author} {\bibfnamefont {M.}~\bibnamefont {Liertzer}}, \bibinfo {author} {\bibfnamefont {S.}~\bibnamefont {Rotter}},\ and\ \bibinfo {author} {\bibfnamefont {H.~E.}\ \bibnamefont {T\"ureci}},\ }\bibfield  {title} {\emph {\bibinfo {title} {Route from spontaneous decay to complex multimode dynamics in cavity qed}},\ }\href {https://doi.org/10.1103/PhysRevA.89.033820} {\bibfield  {journal} {\bibinfo  {journal} {Phys. Rev. A}\ }\textbf {\bibinfo {volume} {89}},\ \bibinfo {pages} {033820} (\bibinfo {year} {2014})}\BibitemShut {NoStop}%
\bibitem [{\citenamefont {Dorner}\ and\ \citenamefont {Zoller}(2002)}]{DornerPRA02}%
  \BibitemOpen
  \bibfield  {author} {\bibinfo {author} {\bibfnamefont {U.}~\bibnamefont {Dorner}}\ and\ \bibinfo {author} {\bibfnamefont {P.}~\bibnamefont {Zoller}},\ }\bibfield  {title} {\emph {\bibinfo {title} {{Laser-driven atoms in half-cavities}}},\ }\href {https://doi.org/10.1103/PhysRevA.66.023816} {\bibfield  {journal} {\bibinfo  {journal} {Physical Review A}\ }\textbf {\bibinfo {volume} {66}},\ \bibinfo {pages} {1--20} (\bibinfo {year} {2002})}\BibitemShut {NoStop}%
\bibitem [{\citenamefont {Tufarelli}\ \emph {et~al.}(2013)\citenamefont {Tufarelli}, \citenamefont {Ciccarello},\ and\ \citenamefont {Kim}}]{TufarelliPRA13}%
  \BibitemOpen
  \bibfield  {author} {\bibinfo {author} {\bibfnamefont {T.}~\bibnamefont {Tufarelli}}, \bibinfo {author} {\bibfnamefont {F.}~\bibnamefont {Ciccarello}},\ and\ \bibinfo {author} {\bibfnamefont {M.~S.}\ \bibnamefont {Kim}},\ }\bibfield  {title} {\emph {\bibinfo {title} {{Dynamics of spontaneous emission in a single-end photonic waveguide}}},\ }\href {https://doi.org/10.1103/PhysRevA.87.013820} {\bibfield  {journal} {\bibinfo  {journal} {Physical Review A}\ }\textbf {\bibinfo {volume} {87}},\ \bibinfo {pages} {13820} (\bibinfo {year} {2013})}\BibitemShut {NoStop}%
\bibitem [{\citenamefont {Tufarelli}\ \emph {et~al.}(2014)\citenamefont {Tufarelli}, \citenamefont {Kim},\ and\ \citenamefont {Ciccarello}}]{TufarelliPRA14}%
  \BibitemOpen
  \bibfield  {author} {\bibinfo {author} {\bibfnamefont {T.}~\bibnamefont {Tufarelli}}, \bibinfo {author} {\bibfnamefont {M.~S.}\ \bibnamefont {Kim}},\ and\ \bibinfo {author} {\bibfnamefont {F.}~\bibnamefont {Ciccarello}},\ }\bibfield  {title} {\emph {\bibinfo {title} {{Non-Markovianity of a quantum emitter in front of a mirror}}},\ }\href {https://doi.org/10.1103/PhysRevA.90.012113} {\bibfield  {journal} {\bibinfo  {journal} {Physical Review A}\ }\textbf {\bibinfo {volume} {90}},\ \bibinfo {pages} {12113} (\bibinfo {year} {2014})}\BibitemShut {NoStop}%
\bibitem [{\citenamefont {Guimond}\ \emph {et~al.}(2017)\citenamefont {Guimond}, \citenamefont {Pletyukhov}, \citenamefont {Pichler},\ and\ \citenamefont {Zoller}}]{guimondDelayed2017}%
  \BibitemOpen
  \bibfield  {author} {\bibinfo {author} {\bibfnamefont {P.~O.}\ \bibnamefont {Guimond}}, \bibinfo {author} {\bibfnamefont {M.}~\bibnamefont {Pletyukhov}}, \bibinfo {author} {\bibfnamefont {H.}~\bibnamefont {Pichler}},\ and\ \bibinfo {author} {\bibfnamefont {P.}~\bibnamefont {Zoller}},\ }\bibfield  {title} {\emph {\bibinfo {title} {{Delayed coherent quantum feedback from a scattering theory and a matrix product state perspective}}},\ }\href {https://doi.org/10.1088/2058-9565/aa7f03} {\bibfield  {journal} {\bibinfo  {journal} {Quantum Science and Technology}\ }\textbf {\bibinfo {volume} {2}},\ \bibinfo {pages} {44012} (\bibinfo {year} {2017})},\ \Eprint {https://arxiv.org/abs/1706.07844} {arXiv:1706.07844} \BibitemShut {NoStop}%
\bibitem [{\citenamefont {Wiegand}\ \emph {et~al.}(2020)\citenamefont {Wiegand}, \citenamefont {Rousseaux},\ and\ \citenamefont {Johansson}}]{GoranPRA20}%
  \BibitemOpen
  \bibfield  {author} {\bibinfo {author} {\bibfnamefont {E.}~\bibnamefont {Wiegand}}, \bibinfo {author} {\bibfnamefont {B.}~\bibnamefont {Rousseaux}},\ and\ \bibinfo {author} {\bibfnamefont {G.}~\bibnamefont {Johansson}},\ }\bibfield  {title} {\emph {\bibinfo {title} {Semiclassical analysis of dark-state transient dynamics in waveguide circuit qed}},\ }\href {https://doi.org/10.1103/PhysRevA.101.033801} {\bibfield  {journal} {\bibinfo  {journal} {Phys. Rev. A}\ }\textbf {\bibinfo {volume} {101}},\ \bibinfo {pages} {033801} (\bibinfo {year} {2020})}\BibitemShut {NoStop}%
\bibitem [{\citenamefont {Guo}\ \emph {et~al.}(2017)\citenamefont {Guo}, \citenamefont {Grimsmo}, \citenamefont {Kockum}, \citenamefont {Pletyukhov},\ and\ \citenamefont {Johansson}}]{GuoPRA17}%
  \BibitemOpen
  \bibfield  {author} {\bibinfo {author} {\bibfnamefont {L.}~\bibnamefont {Guo}}, \bibinfo {author} {\bibfnamefont {A.}~\bibnamefont {Grimsmo}}, \bibinfo {author} {\bibfnamefont {A.~F.}\ \bibnamefont {Kockum}}, \bibinfo {author} {\bibfnamefont {M.}~\bibnamefont {Pletyukhov}},\ and\ \bibinfo {author} {\bibfnamefont {G.}~\bibnamefont {Johansson}},\ }\bibfield  {title} {\emph {\bibinfo {title} {{Giant acoustic atom: A single quantum system with a deterministic time delay}}},\ }\href {https://doi.org/10.1103/PhysRevA.95.053821} {\bibfield  {journal} {\bibinfo  {journal} {Physical Review A}\ }\textbf {\bibinfo {volume} {95}},\ \bibinfo {pages} {53821} (\bibinfo {year} {2017})},\ \Eprint {https://arxiv.org/abs/1612.00865} {arXiv:1612.00865} \BibitemShut {NoStop}%
\bibitem [{\citenamefont {Gonzalez-Ballestero}\ \emph {et~al.}(2016)\citenamefont {Gonzalez-Ballestero}, \citenamefont {Moreno}, \citenamefont {Garcia-Vidal},\ and\ \citenamefont {Gonzalez-Tudela}}]{Gonzalez-BallesteroPRA16}%
  \BibitemOpen
  \bibfield  {author} {\bibinfo {author} {\bibfnamefont {C.}~\bibnamefont {Gonzalez-Ballestero}}, \bibinfo {author} {\bibfnamefont {E.}~\bibnamefont {Moreno}}, \bibinfo {author} {\bibfnamefont {F.~J.}\ \bibnamefont {Garcia-Vidal}},\ and\ \bibinfo {author} {\bibfnamefont {A.}~\bibnamefont {Gonzalez-Tudela}},\ }\bibfield  {title} {\emph {\bibinfo {title} {{Nonreciprocal few-photon routing schemes based on chiral waveguide-emitter couplings}}},\ }\href {https://doi.org/10.1103/PhysRevA.94.063817} {\bibfield  {journal} {\bibinfo  {journal} {Physical Review A}\ }\textbf {\bibinfo {volume} {94}},\ \bibinfo {pages} {63817} (\bibinfo {year} {2016})},\ \Eprint {https://arxiv.org/abs/1608.04928} {arXiv:1608.04928} \BibitemShut {NoStop}%
\bibitem [{\citenamefont {Fang}\ \emph {et~al.}(2018)\citenamefont {Fang}, \citenamefont {Ciccarello},\ and\ \citenamefont {Baranger}}]{FangNJP18}%
  \BibitemOpen
  \bibfield  {author} {\bibinfo {author} {\bibfnamefont {Y.~L.~L.}\ \bibnamefont {Fang}}, \bibinfo {author} {\bibfnamefont {F.}~\bibnamefont {Ciccarello}},\ and\ \bibinfo {author} {\bibfnamefont {H.~U.}\ \bibnamefont {Baranger}},\ }\bibfield  {title} {\emph {\bibinfo {title} {{Non-Markovian dynamics of a qubit due to single-photon scattering in a waveguide}}},\ }\href {https://doi.org/10.1088/1367-2630/aaba5d} {\bibfield  {journal} {\bibinfo  {journal} {New Journal of Physics}\ }\textbf {\bibinfo {volume} {20}},\ \bibinfo {pages} {43035} (\bibinfo {year} {2018})}\BibitemShut {NoStop}%
\bibitem [{SM()}]{SM}%
  \BibitemOpen
  \href@noop {} {}\bibinfo {howpublished} {See Supplemental Material at xxx for technical details.}\BibitemShut {Stop}%
\bibitem [{\citenamefont {Viviescas}\ and\ \citenamefont {Hackenbroich}(2003)}]{viviescas2003field}%
  \BibitemOpen
  \bibfield  {author} {\bibinfo {author} {\bibfnamefont {C.}~\bibnamefont {Viviescas}}\ and\ \bibinfo {author} {\bibfnamefont {G.}~\bibnamefont {Hackenbroich}},\ }\bibfield  {title} {\emph {\bibinfo {title} {Field quantization for open optical cavities}},\ }\href@noop {} {\bibfield  {journal} {\bibinfo  {journal} {Physical Review A}\ }\textbf {\bibinfo {volume} {67}},\ \bibinfo {pages} {013805} (\bibinfo {year} {2003})}\BibitemShut {NoStop}%
\bibitem [{\citenamefont {Lentrodt}\ and\ \citenamefont {Evers}(2020)}]{lentrodt2020ab}%
  \BibitemOpen
  \bibfield  {author} {\bibinfo {author} {\bibfnamefont {D.}~\bibnamefont {Lentrodt}}\ and\ \bibinfo {author} {\bibfnamefont {J.}~\bibnamefont {Evers}},\ }\bibfield  {title} {\emph {\bibinfo {title} {Ab initio few-mode theory for quantum potential scattering problems}},\ }\href@noop {} {\bibfield  {journal} {\bibinfo  {journal} {Physical Review X}\ }\textbf {\bibinfo {volume} {10}},\ \bibinfo {pages} {011008} (\bibinfo {year} {2020})}\BibitemShut {NoStop}%
\bibitem [{\citenamefont {Hoi}\ \emph {et~al.}(2015)\citenamefont {Hoi}, \citenamefont {Kockum}, \citenamefont {Tornberg}, \citenamefont {Pourkabirian}, \citenamefont {Johansson}, \citenamefont {Delsing},\ and\ \citenamefont {Wilson}}]{HoiNatPhy15}%
  \BibitemOpen
  \bibfield  {author} {\bibinfo {author} {\bibfnamefont {I.~C.}\ \bibnamefont {Hoi}}, \bibinfo {author} {\bibfnamefont {A.~F.}\ \bibnamefont {Kockum}}, \bibinfo {author} {\bibfnamefont {L.}~\bibnamefont {Tornberg}}, \bibinfo {author} {\bibfnamefont {A.}~\bibnamefont {Pourkabirian}}, \bibinfo {author} {\bibfnamefont {G.}~\bibnamefont {Johansson}}, \bibinfo {author} {\bibfnamefont {P.}~\bibnamefont {Delsing}},\ and\ \bibinfo {author} {\bibfnamefont {C.~M.}\ \bibnamefont {Wilson}},\ }\bibfield  {title} {\emph {\bibinfo {title} {{Probing the quantum vacuum with an artificial atom in front of a mirror}}},\ }\href {https://doi.org/10.1038/nphys3484} {\bibfield  {journal} {\bibinfo  {journal} {Nature Physics}\ }\textbf {\bibinfo {volume} {11}},\ \bibinfo {pages} {1045--1049} (\bibinfo {year} {2015})}\BibitemShut {NoStop}%
\bibitem [{\citenamefont {Ask}\ \emph {et~al.}(2019)\citenamefont {Ask}, \citenamefont {Ekstr\"om}, \citenamefont {Delsing},\ and\ \citenamefont {Johansson}}]{AskPRA19}%
  \BibitemOpen
  \bibfield  {author} {\bibinfo {author} {\bibfnamefont {A.}~\bibnamefont {Ask}}, \bibinfo {author} {\bibfnamefont {M.}~\bibnamefont {Ekstr\"om}}, \bibinfo {author} {\bibfnamefont {P.}~\bibnamefont {Delsing}},\ and\ \bibinfo {author} {\bibfnamefont {G.}~\bibnamefont {Johansson}},\ }\bibfield  {title} {\emph {\bibinfo {title} {Cavity-free vacuum-rabi splitting in circuit quantum acoustodynamics}},\ }\href {https://doi.org/10.1103/PhysRevA.99.013840} {\bibfield  {journal} {\bibinfo  {journal} {Phys. Rev. A}\ }\textbf {\bibinfo {volume} {99}},\ \bibinfo {pages} {013840} (\bibinfo {year} {2019})}\BibitemShut {NoStop}%
\bibitem [{\citenamefont {Lindberg}\ and\ \citenamefont {Savage}(1988)}]{lindberg1988steady}%
  \BibitemOpen
  \bibfield  {author} {\bibinfo {author} {\bibfnamefont {M.}~\bibnamefont {Lindberg}}\ and\ \bibinfo {author} {\bibfnamefont {C.~M.}\ \bibnamefont {Savage}},\ }\bibfield  {title} {\emph {\bibinfo {title} {{Steady-state two-level atomic population inversion via a quantized cavity field}}},\ }\href {https://journals.aps.org/pra/abstract/10.1103/PhysRevA.38.5182} {\bibfield  {journal} {\bibinfo  {journal} {Physical Review A}\ }\textbf {\bibinfo {volume} {38}},\ \bibinfo {pages} {5182} (\bibinfo {year} {1988})}\BibitemShut {NoStop}%
\bibitem [{\citenamefont {Chang}\ \emph {et~al.}(2012)\citenamefont {Chang}, \citenamefont {Jiang}, \citenamefont {Gorshkov},\ and\ \citenamefont {Kimble}}]{ChangNJP12}%
  \BibitemOpen
  \bibfield  {author} {\bibinfo {author} {\bibfnamefont {D.~E.}\ \bibnamefont {Chang}}, \bibinfo {author} {\bibfnamefont {L.}~\bibnamefont {Jiang}}, \bibinfo {author} {\bibfnamefont {A.~V.}\ \bibnamefont {Gorshkov}},\ and\ \bibinfo {author} {\bibfnamefont {H.~J.}\ \bibnamefont {Kimble}},\ }\bibfield  {title} {\emph {\bibinfo {title} {{Cavity QED with atomic mirrors}}},\ }\href {https://doi.org/10.1088/1367-2630/14/6/063003} {\bibfield  {journal} {\bibinfo  {journal} {New Journal of Physics}\ }\textbf {\bibinfo {volume} {14}},\ \bibinfo {pages} {63003} (\bibinfo {year} {2012})}\BibitemShut {NoStop}%
\bibitem [{\citenamefont {Guimond}\ \emph {et~al.}(2016)\citenamefont {Guimond}, \citenamefont {Roulet}, \citenamefont {Le},\ and\ \citenamefont {Scarani}}]{GuimondPRA16}%
  \BibitemOpen
  \bibfield  {author} {\bibinfo {author} {\bibfnamefont {P.~O.}\ \bibnamefont {Guimond}}, \bibinfo {author} {\bibfnamefont {A.}~\bibnamefont {Roulet}}, \bibinfo {author} {\bibfnamefont {H.~N.}\ \bibnamefont {Le}},\ and\ \bibinfo {author} {\bibfnamefont {V.}~\bibnamefont {Scarani}},\ }\bibfield  {title} {\emph {\bibinfo {title} {{Rabi oscillation in a quantum cavity: Markovian and non-Markovian dynamics}}},\ }\href {https://doi.org/10.1103/PhysRevA.93.023808} {\bibfield  {journal} {\bibinfo  {journal} {Physical Review A}\ }\textbf {\bibinfo {volume} {93}},\ \bibinfo {pages} {23808} (\bibinfo {year} {2016})},\ \Eprint {https://arxiv.org/abs/1505.07908} {arXiv:1505.07908} \BibitemShut {NoStop}%
\bibitem [{\citenamefont {Wiegand}\ \emph {et~al.}(2021)\citenamefont {Wiegand}, \citenamefont {Rousseaux},\ and\ \citenamefont {Johansson}}]{GoranPRR21}%
  \BibitemOpen
  \bibfield  {author} {\bibinfo {author} {\bibfnamefont {E.}~\bibnamefont {Wiegand}}, \bibinfo {author} {\bibfnamefont {B.}~\bibnamefont {Rousseaux}},\ and\ \bibinfo {author} {\bibfnamefont {G.}~\bibnamefont {Johansson}},\ }\bibfield  {title} {\emph {\bibinfo {title} {Transmon in a semi-infinite high-impedance transmission line: Appearance of cavity modes and rabi oscillations}},\ }\href {https://doi.org/10.1103/PhysRevResearch.3.023003} {\bibfield  {journal} {\bibinfo  {journal} {Phys. Rev. Res.}\ }\textbf {\bibinfo {volume} {3}},\ \bibinfo {pages} {023003} (\bibinfo {year} {2021})}\BibitemShut {NoStop}%
\bibitem [{\citenamefont {Arranz~Regidor}\ and\ \citenamefont {Hughes}(2023)}]{Hughes3qubits}%
  \BibitemOpen
  \bibfield  {author} {\bibinfo {author} {\bibfnamefont {S.}~\bibnamefont {Arranz~Regidor}}\ and\ \bibinfo {author} {\bibfnamefont {S.}~\bibnamefont {Hughes}},\ }\bibfield  {title} {\emph {\bibinfo {title} {Probing dressed states and quantum nonlinearities in a strongly coupled three-qubit waveguide system under optical pumping}},\ }\href {https://doi.org/10.1103/PhysRevA.108.033719} {\bibfield  {journal} {\bibinfo  {journal} {Phys. Rev. A}\ }\textbf {\bibinfo {volume} {108}},\ \bibinfo {pages} {033719} (\bibinfo {year} {2023})}\BibitemShut {NoStop}%
\bibitem [{\citenamefont {Shen}\ and\ \citenamefont {Fan}(2005)}]{Shen2005}%
  \BibitemOpen
  \bibfield  {author} {\bibinfo {author} {\bibfnamefont {J.~T.}\ \bibnamefont {Shen}}\ and\ \bibinfo {author} {\bibfnamefont {S.}~\bibnamefont {Fan}},\ }\bibfield  {title} {\emph {\bibinfo {title} {{Coherent photon transport from spontaneous emission in one-dimensional waveguides}}},\ }\href {https://doi.org/10.1364/ol.30.002001} {\bibfield  {journal} {\bibinfo  {journal} {Optics Letters}\ }\textbf {\bibinfo {volume} {30}},\ \bibinfo {pages} {2001} (\bibinfo {year} {2005})}\BibitemShut {NoStop}%
\bibitem [{\citenamefont {Tamascelli}\ \emph {et~al.}(2018)\citenamefont {Tamascelli}, \citenamefont {Smirne}, \citenamefont {Huelga},\ and\ \citenamefont {Plenio}}]{TamaPRL18}%
  \BibitemOpen
  \bibfield  {author} {\bibinfo {author} {\bibfnamefont {D.}~\bibnamefont {Tamascelli}}, \bibinfo {author} {\bibfnamefont {A.}~\bibnamefont {Smirne}}, \bibinfo {author} {\bibfnamefont {S.~F.}\ \bibnamefont {Huelga}},\ and\ \bibinfo {author} {\bibfnamefont {M.~B.}\ \bibnamefont {Plenio}},\ }\bibfield  {title} {\emph {\bibinfo {title} {Nonperturbative treatment of non-markovian dynamics of open quantum systems}},\ }\href {https://doi.org/10.1103/PhysRevLett.120.030402} {\bibfield  {journal} {\bibinfo  {journal} {Phys. Rev. Lett.}\ }\textbf {\bibinfo {volume} {120}},\ \bibinfo {pages} {030402} (\bibinfo {year} {2018})}\BibitemShut {NoStop}%
\bibitem [{\citenamefont {Campbell}\ \emph {et~al.}(2018)\citenamefont {Campbell}, \citenamefont {Ciccarello}, \citenamefont {Palma},\ and\ \citenamefont {Vacchini}}]{CampbellPRA18}%
  \BibitemOpen
  \bibfield  {author} {\bibinfo {author} {\bibfnamefont {S.}~\bibnamefont {Campbell}}, \bibinfo {author} {\bibfnamefont {F.}~\bibnamefont {Ciccarello}}, \bibinfo {author} {\bibfnamefont {G.~M.}\ \bibnamefont {Palma}},\ and\ \bibinfo {author} {\bibfnamefont {B.}~\bibnamefont {Vacchini}},\ }\bibfield  {title} {\emph {\bibinfo {title} {System-environment correlations and markovian embedding of quantum non-markovian dynamics}},\ }\href {https://doi.org/10.1103/PhysRevA.98.012142} {\bibfield  {journal} {\bibinfo  {journal} {Phys. Rev. A}\ }\textbf {\bibinfo {volume} {98}},\ \bibinfo {pages} {012142} (\bibinfo {year} {2018})}\BibitemShut {NoStop}%
\bibitem [{\citenamefont {Trivedi}\ \emph {et~al.}(2021{\natexlab{b}})\citenamefont {Trivedi}, \citenamefont {Malz},\ and\ \citenamefont {Cirac}}]{TrivediPRL21}%
  \BibitemOpen
  \bibfield  {author} {\bibinfo {author} {\bibfnamefont {R.}~\bibnamefont {Trivedi}}, \bibinfo {author} {\bibfnamefont {D.}~\bibnamefont {Malz}},\ and\ \bibinfo {author} {\bibfnamefont {J.~I.}\ \bibnamefont {Cirac}},\ }\bibfield  {title} {\emph {\bibinfo {title} {Convergence guarantees for discrete mode approximations to non-markovian quantum baths}},\ }\href {https://doi.org/10.1103/PhysRevLett.127.250404} {\bibfield  {journal} {\bibinfo  {journal} {Phys. Rev. Lett.}\ }\textbf {\bibinfo {volume} {127}},\ \bibinfo {pages} {250404} (\bibinfo {year} {2021}{\natexlab{b}})}\BibitemShut {NoStop}%
\bibitem [{\citenamefont {N\"u\ss{}eler}\ \emph {et~al.}(2022)\citenamefont {N\"u\ss{}eler}, \citenamefont {Tamascelli}, \citenamefont {Smirne}, \citenamefont {Lim}, \citenamefont {Huelga},\ and\ \citenamefont {Plenio}}]{TamaPRL22}%
  \BibitemOpen
  \bibfield  {author} {\bibinfo {author} {\bibfnamefont {A.}~\bibnamefont {N\"u\ss{}eler}}, \bibinfo {author} {\bibfnamefont {D.}~\bibnamefont {Tamascelli}}, \bibinfo {author} {\bibfnamefont {A.}~\bibnamefont {Smirne}}, \bibinfo {author} {\bibfnamefont {J.}~\bibnamefont {Lim}}, \bibinfo {author} {\bibfnamefont {S.~F.}\ \bibnamefont {Huelga}},\ and\ \bibinfo {author} {\bibfnamefont {M.~B.}\ \bibnamefont {Plenio}},\ }\bibfield  {title} {\emph {\bibinfo {title} {Fingerprint and universal markovian closure of structured bosonic environments}},\ }\href {https://doi.org/10.1103/PhysRevLett.129.140604} {\bibfield  {journal} {\bibinfo  {journal} {Phys. Rev. Lett.}\ }\textbf {\bibinfo {volume} {129}},\ \bibinfo {pages} {140604} (\bibinfo {year} {2022})}\BibitemShut {NoStop}%
\bibitem [{\citenamefont {Johansson}\ \emph {et~al.}(2012)\citenamefont {Johansson}, \citenamefont {Nation},\ and\ \citenamefont {Nori}}]{JOHANSSON20121760}%
  \BibitemOpen
  \bibfield  {author} {\bibinfo {author} {\bibfnamefont {J.}~\bibnamefont {Johansson}}, \bibinfo {author} {\bibfnamefont {P.}~\bibnamefont {Nation}},\ and\ \bibinfo {author} {\bibfnamefont {F.}~\bibnamefont {Nori}},\ }\bibfield  {title} {\emph {\bibinfo {title} {Qutip: An open-source python framework for the dynamics of open quantum systems}},\ }\href {https://doi.org/https://doi.org/10.1016/j.cpc.2012.02.021} {\bibfield  {journal} {\bibinfo  {journal} {Computer Physics Communications}\ }\textbf {\bibinfo {volume} {183}},\ \bibinfo {pages} {1760--1772} (\bibinfo {year} {2012})}\BibitemShut {NoStop}%
\bibitem [{\citenamefont {Suess}\ and\ \citenamefont {Holzäpfel}(2017)}]{suess2017mpnum}%
  \BibitemOpen
  \bibfield  {author} {\bibinfo {author} {\bibfnamefont {D.}~\bibnamefont {Suess}}\ and\ \bibinfo {author} {\bibfnamefont {M.}~\bibnamefont {Holzäpfel}},\ }\bibfield  {title} {\emph {\bibinfo {title} {mpnum: A matrix product representation library for python}},\ }\href {https://doi.org/10.21105/joss.00465} {\bibfield  {journal} {\bibinfo  {journal} {Journal of Open Source Software}\ }\textbf {\bibinfo {volume} {2}},\ \bibinfo {pages} {465} (\bibinfo {year} {2017})}\BibitemShut {NoStop}%
\end{thebibliography}%


\begin{thebibliography}{14}%
\makeatletter
\providecommand \@ifxundefined [1]{%
 \@ifx{#1\undefined}
}%
\providecommand \@ifnum [1]{%
 \ifnum #1\expandafter \@firstoftwo
 \else \expandafter \@secondoftwo
 \fi
}%
\providecommand \@ifx [1]{%
 \ifx #1\expandafter \@firstoftwo
 \else \expandafter \@secondoftwo
 \fi
}%
\providecommand \natexlab [1]{#1}%
\providecommand \enquote  [1]{``#1''}%
\providecommand \bibnamefont  [1]{#1}%
\providecommand \bibfnamefont [1]{#1}%
\providecommand \citenamefont [1]{#1}%
\providecommand \href@noop [0]{\@secondoftwo}%
\providecommand \href [0]{\begingroup \@sanitize@url \@href}%
\providecommand \@href[1]{\@@startlink{#1}\@@href}%
\providecommand \@@href[1]{\endgroup#1\@@endlink}%
\providecommand \@sanitize@url [0]{\catcode `\\12\catcode `\$12\catcode `\&12\catcode `\#12\catcode `\^12\catcode `\_12\catcode `\%12\relax}%
\providecommand \@@startlink[1]{}%
\providecommand \@@endlink[0]{}%
\providecommand \url  [0]{\begingroup\@sanitize@url \@url }%
\providecommand \@url [1]{\endgroup\@href {#1}{\urlprefix }}%
\providecommand \urlprefix  [0]{URL }%
\providecommand \Eprint [0]{\href }%
\providecommand \doibase [0]{https://doi.org/}%
\providecommand \selectlanguage [0]{\@gobble}%
\providecommand \bibinfo  [0]{\@secondoftwo}%
\providecommand \bibfield  [0]{\@secondoftwo}%
\providecommand \translation [1]{[#1]}%
\providecommand \BibitemOpen [0]{}%
\providecommand \bibitemStop [0]{}%
\providecommand \bibitemNoStop [0]{.\EOS\space}%
\providecommand \EOS [0]{\spacefactor3000\relax}%
\providecommand \BibitemShut  [1]{\csname bibitem#1\endcsname}%
\let\auto@bib@innerbib\@empty
\bibitem [{\citenamefont {Dorner}\ and\ \citenamefont {Zoller}(2002)}]{DornerPRA02}%
  \BibitemOpen
  \bibfield  {author} {\bibinfo {author} {\bibfnamefont {U.}~\bibnamefont {Dorner}}\ and\ \bibinfo {author} {\bibfnamefont {P.}~\bibnamefont {Zoller}},\ }\bibfield  {title} {\emph {\bibinfo {title} {{Laser-driven atoms in half-cavities}}},\ }\href {https://doi.org/10.1103/PhysRevA.66.023816} {\bibfield  {journal} {\bibinfo  {journal} {Physical Review A}\ }\textbf {\bibinfo {volume} {66}},\ \bibinfo {pages} {1--20} (\bibinfo {year} {2002})}\BibitemShut {NoStop}%
\bibitem [{\citenamefont {Tufarelli}\ \emph {et~al.}(2013)\citenamefont {Tufarelli}, \citenamefont {Ciccarello},\ and\ \citenamefont {Kim}}]{TufarelliPRA13}%
  \BibitemOpen
  \bibfield  {author} {\bibinfo {author} {\bibfnamefont {T.}~\bibnamefont {Tufarelli}}, \bibinfo {author} {\bibfnamefont {F.}~\bibnamefont {Ciccarello}},\ and\ \bibinfo {author} {\bibfnamefont {M.~S.}\ \bibnamefont {Kim}},\ }\bibfield  {title} {\emph {\bibinfo {title} {{Dynamics of spontaneous emission in a single-end photonic waveguide}}},\ }\href {https://doi.org/10.1103/PhysRevA.87.013820} {\bibfield  {journal} {\bibinfo  {journal} {Physical Review A}\ }\textbf {\bibinfo {volume} {87}},\ \bibinfo {pages} {13820} (\bibinfo {year} {2013})}\BibitemShut {NoStop}%
\bibitem [{\citenamefont {Pichler}\ and\ \citenamefont {Zoller}(2016)}]{pichlerPhotonic2016}%
  \BibitemOpen
  \bibfield  {author} {\bibinfo {author} {\bibfnamefont {H.}~\bibnamefont {Pichler}}\ and\ \bibinfo {author} {\bibfnamefont {P.}~\bibnamefont {Zoller}},\ }\bibfield  {title} {\emph {\bibinfo {title} {{Photonic Circuits with Time Delays and Quantum Feedback}}},\ }\href {https://doi.org/10.1103/PhysRevLett.116.093601} {\bibfield  {journal} {\bibinfo  {journal} {Physical Review Letters}\ }\textbf {\bibinfo {volume} {116}},\ \bibinfo {pages} {93601} (\bibinfo {year} {2016})}\BibitemShut {NoStop}%
\bibitem [{\citenamefont {Fang}\ \emph {et~al.}(2018)\citenamefont {Fang}, \citenamefont {Ciccarello},\ and\ \citenamefont {Baranger}}]{FangNJP18}%
  \BibitemOpen
  \bibfield  {author} {\bibinfo {author} {\bibfnamefont {Y.~L.~L.}\ \bibnamefont {Fang}}, \bibinfo {author} {\bibfnamefont {F.}~\bibnamefont {Ciccarello}},\ and\ \bibinfo {author} {\bibfnamefont {H.~U.}\ \bibnamefont {Baranger}},\ }\bibfield  {title} {\emph {\bibinfo {title} {{Non-Markovian dynamics of a qubit due to single-photon scattering in a waveguide}}},\ }\href {https://doi.org/10.1088/1367-2630/aaba5d} {\bibfield  {journal} {\bibinfo  {journal} {New Journal of Physics}\ }\textbf {\bibinfo {volume} {20}},\ \bibinfo {pages} {43035} (\bibinfo {year} {2018})}\BibitemShut {NoStop}%
\bibitem [{\citenamefont {Cilluffo}\ \emph {et~al.}(2020)\citenamefont {Cilluffo}, \citenamefont {Carollo}, \citenamefont {Lorenzo}, \citenamefont {Gross}, \citenamefont {Palma},\ and\ \citenamefont {Ciccarello}}]{DarioTB}%
  \BibitemOpen
  \bibfield  {author} {\bibinfo {author} {\bibfnamefont {D.}~\bibnamefont {Cilluffo}}, \bibinfo {author} {\bibfnamefont {A.}~\bibnamefont {Carollo}}, \bibinfo {author} {\bibfnamefont {S.}~\bibnamefont {Lorenzo}}, \bibinfo {author} {\bibfnamefont {J.~A.}\ \bibnamefont {Gross}}, \bibinfo {author} {\bibfnamefont {G.~M.}\ \bibnamefont {Palma}},\ and\ \bibinfo {author} {\bibfnamefont {F.}~\bibnamefont {Ciccarello}},\ }\bibfield  {title} {\emph {\bibinfo {title} {Collisional picture of quantum optics with giant emitters}},\ }\href {https://doi.org/10.1103/PhysRevResearch.2.043070} {\bibfield  {journal} {\bibinfo  {journal} {Phys. Rev. Research}\ }\textbf {\bibinfo {volume} {2}},\ \bibinfo {pages} {043070} (\bibinfo {year} {2020})}\BibitemShut {NoStop}%
\bibitem [{\citenamefont {Ciccarello}\ \emph {et~al.}(2022)\citenamefont {Ciccarello}, \citenamefont {Lorenzo}, \citenamefont {Giovannetti},\ and\ \citenamefont {Palma}}]{Ciccarello_2022}%
  \BibitemOpen
  \bibfield  {author} {\bibinfo {author} {\bibfnamefont {F.}~\bibnamefont {Ciccarello}}, \bibinfo {author} {\bibfnamefont {S.}~\bibnamefont {Lorenzo}}, \bibinfo {author} {\bibfnamefont {V.}~\bibnamefont {Giovannetti}},\ and\ \bibinfo {author} {\bibfnamefont {G.~M.}\ \bibnamefont {Palma}},\ }\bibfield  {title} {\emph {\bibinfo {title} {Quantum collision models: Open system dynamics from repeated interactions}},\ }\href {https://doi.org/10.1016/j.physrep.2022.01.001} {\bibfield  {journal} {\bibinfo  {journal} {Physics Reports}\ }\textbf {\bibinfo {volume} {954}},\ \bibinfo {pages} {1–70} (\bibinfo {year} {2022})}\BibitemShut {NoStop}%
\bibitem [{\citenamefont {Schollw{\"{o}}ck}(2011)}]{SchollwockAnnPhys11}%
  \BibitemOpen
  \bibfield  {author} {\bibinfo {author} {\bibfnamefont {U.}~\bibnamefont {Schollw{\"{o}}ck}},\ }\bibfield  {title} {\emph {\bibinfo {title} {{The density-matrix renormalization group in the age of matrix product states}}},\ }\href {https://doi.org/10.1016/j.aop.2010.09.012} {\bibfield  {journal} {\bibinfo  {journal} {Annals of Physics}\ }\textbf {\bibinfo {volume} {326}},\ \bibinfo {pages} {96--192} (\bibinfo {year} {2011})},\ \Eprint {https://arxiv.org/abs/1008.3477} {arXiv:1008.3477} \BibitemShut {NoStop}%
\bibitem [{\citenamefont {Vidal}(2003)}]{VidalPRL03}%
  \BibitemOpen
  \bibfield  {author} {\bibinfo {author} {\bibfnamefont {G.}~\bibnamefont {Vidal}},\ }\bibfield  {title} {\emph {\bibinfo {title} {{Efficient classical simulation of slightly entangled quantum computations}}},\ }\href {https://doi.org/10.1103/PhysRevLett.91.147902} {\bibfield  {journal} {\bibinfo  {journal} {Physical Review Letters}\ }\textbf {\bibinfo {volume} {91}},\ \bibinfo {pages} {147902} (\bibinfo {year} {2003})},\ \Eprint {https://arxiv.org/abs/0301063} {arXiv:0301063 [quant-ph]} \BibitemShut {NoStop}%
\bibitem [{\citenamefont {Schachenmayer}\ \emph {et~al.}(2010)\citenamefont {Schachenmayer}, \citenamefont {Lesanovsky}, \citenamefont {Micheli},\ and\ \citenamefont {Daley}}]{Schachenmayer_2010}%
  \BibitemOpen
  \bibfield  {author} {\bibinfo {author} {\bibfnamefont {J.}~\bibnamefont {Schachenmayer}}, \bibinfo {author} {\bibfnamefont {I.}~\bibnamefont {Lesanovsky}}, \bibinfo {author} {\bibfnamefont {A.}~\bibnamefont {Micheli}},\ and\ \bibinfo {author} {\bibfnamefont {A.~J.}\ \bibnamefont {Daley}},\ }\bibfield  {title} {\emph {\bibinfo {title} {Dynamical crystal creation with polar molecules or rydberg atoms in optical lattices}},\ }\href {https://doi.org/10.1088/1367-2630/12/10/103044} {\bibfield  {journal} {\bibinfo  {journal} {New Journal of Physics}\ }\textbf {\bibinfo {volume} {12}},\ \bibinfo {pages} {103044} (\bibinfo {year} {2010})}\BibitemShut {NoStop}%
\bibitem [{\citenamefont {Cuevas}\ \emph {et~al.}(2013)\citenamefont {Cuevas}, \citenamefont {Schuch}, \citenamefont {Pérez-García},\ and\ \citenamefont {Ignacio~Cirac}}]{Cuevas_2013}%
  \BibitemOpen
  \bibfield  {author} {\bibinfo {author} {\bibfnamefont {G.~D.~l.}\ \bibnamefont {Cuevas}}, \bibinfo {author} {\bibfnamefont {N.}~\bibnamefont {Schuch}}, \bibinfo {author} {\bibfnamefont {D.}~\bibnamefont {Pérez-García}},\ and\ \bibinfo {author} {\bibfnamefont {J.}~\bibnamefont {Ignacio~Cirac}},\ }\bibfield  {title} {\emph {\bibinfo {title} {Purifications of multipartite states: limitations and constructive methods}},\ }\href {https://doi.org/10.1088/1367-2630/15/12/123021} {\bibfield  {journal} {\bibinfo  {journal} {New Journal of Physics}\ }\textbf {\bibinfo {volume} {15}},\ \bibinfo {pages} {123021} (\bibinfo {year} {2013})}\BibitemShut {NoStop}%
\bibitem [{\citenamefont {Zheng}\ \emph {et~al.}(2010)\citenamefont {Zheng}, \citenamefont {Gauthier},\ and\ \citenamefont {Baranger}}]{Zheng2010}%
  \BibitemOpen
  \bibfield  {author} {\bibinfo {author} {\bibfnamefont {H.}~\bibnamefont {Zheng}}, \bibinfo {author} {\bibfnamefont {D.~J.}\ \bibnamefont {Gauthier}},\ and\ \bibinfo {author} {\bibfnamefont {H.~U.}\ \bibnamefont {Baranger}},\ }\bibfield  {title} {\emph {\bibinfo {title} {{Waveguide QED: Many-body bound-state effects in coherent and Fock-state scattering from a two-level system}}},\ }\bibfield  {journal} {\bibinfo  {journal} {Physical Review A}\ }\textbf {\bibinfo {volume} {82}},\ \href {https://doi.org/10.1103/PhysRevA.82.063816} {10.1103/PhysRevA.82.063816} (\bibinfo {year} {2010}),\ \Eprint {https://arxiv.org/abs/1009.5325} {arXiv:1009.5325} \BibitemShut {NoStop}%
\bibitem [{\citenamefont {M{\o}lmer}\ \emph {et~al.}(1993)\citenamefont {M{\o}lmer}, \citenamefont {Castin},\ and\ \citenamefont {Dalibard}}]{molmer1993monte}%
  \BibitemOpen
  \bibfield  {author} {\bibinfo {author} {\bibfnamefont {K.}~\bibnamefont {M{\o}lmer}}, \bibinfo {author} {\bibfnamefont {Y.}~\bibnamefont {Castin}},\ and\ \bibinfo {author} {\bibfnamefont {J.}~\bibnamefont {Dalibard}},\ }\bibfield  {title} {\emph {\bibinfo {title} {Monte carlo wave-function method in quantum optics}},\ }\href {https://doi.org/10.1364/JOSAB.10.000524} {\bibfield  {journal} {\bibinfo  {journal} {J. Opt. Soc. Am. B}\ }\textbf {\bibinfo {volume} {10}},\ \bibinfo {pages} {524--538} (\bibinfo {year} {1993})}\BibitemShut {NoStop}%
\bibitem [{\citenamefont {Gardiner}\ and\ \citenamefont {Collett}(1985)}]{gardiner1985input}%
  \BibitemOpen
  \bibfield  {author} {\bibinfo {author} {\bibfnamefont {C.~W.}\ \bibnamefont {Gardiner}}\ and\ \bibinfo {author} {\bibfnamefont {M.~J.}\ \bibnamefont {Collett}},\ }\bibfield  {title} {\emph {\bibinfo {title} {Input and output in damped quantum systems: Quantum stochastic differential equations and the master equation}},\ }\href {https://doi.org/10.1103/PhysRevA.31.3761} {\bibfield  {journal} {\bibinfo  {journal} {Phys. Rev. A}\ }\textbf {\bibinfo {volume} {31}},\ \bibinfo {pages} {3761--3774} (\bibinfo {year} {1985})}\BibitemShut {NoStop}%
\bibitem [{\citenamefont {Sinha}\ \emph {et~al.}(2020)\citenamefont {Sinha}, \citenamefont {Meystre}, \citenamefont {Goldschmidt}, \citenamefont {Fatemi}, \citenamefont {Rolston},\ and\ \citenamefont {Solano}}]{SinhaPRL20}%
  \BibitemOpen
  \bibfield  {author} {\bibinfo {author} {\bibfnamefont {K.}~\bibnamefont {Sinha}}, \bibinfo {author} {\bibfnamefont {P.}~\bibnamefont {Meystre}}, \bibinfo {author} {\bibfnamefont {E.~A.}\ \bibnamefont {Goldschmidt}}, \bibinfo {author} {\bibfnamefont {F.~K.}\ \bibnamefont {Fatemi}}, \bibinfo {author} {\bibfnamefont {S.~L.}\ \bibnamefont {Rolston}},\ and\ \bibinfo {author} {\bibfnamefont {P.}~\bibnamefont {Solano}},\ }\bibfield  {title} {\emph {\bibinfo {title} {Non-markovian collective emission from macroscopically separated emitters}},\ }\href {https://doi.org/10.1103/PhysRevLett.124.043603} {\bibfield  {journal} {\bibinfo  {journal} {Phys. Rev. Lett.}\ }\textbf {\bibinfo {volume} {124}},\ \bibinfo {pages} {043603} (\bibinfo {year} {2020})}\BibitemShut {NoStop}%
\end{thebibliography}%
\end{document}